\documentclass[%
 reprint,
superscriptaddress,
amsmath,amssymb,
aps,
prb,
floatfix,
longbibliography
]{revtex4-2}

\usepackage[english]{babel}

\usepackage{amsmath,amssymb,amsthm}
\usepackage{tikz}

\usepackage{pgfplots}
\pgfplotsset{width=10cm,compat=1.9}

\usepackage{physics}
\usepackage{graphicx}
\usepackage[colorlinks=true, allcolors=blue]{hyperref}

\definecolor{blau}{RGB}{50,50,150}

\begin{document}

\title{Anomalous Structures of Fisher Zero Lines in Biorthogonal Dynamical Quantum Phase Transitions}

\author{Wen-Hui Nie}
\affiliation{School of Physics, Dalian University of Technology, Dalian 116024, China}

\author{Chong Li}
\affiliation{School of Physics, Dalian University of Technology, Dalian 116024, China}

\author{Lin-Cheng Wang}
\email{wanglc@dlut.edu.cn}
\affiliation{School of Physics, Dalian University of Technology, Dalian 116024, China}

\date{\today}

\begin{abstract}
Fisher zero lines of the Loschmidt amplitude in non-Hermitian quenches can exhibit anomalous structures in the complex time plane, such as divergence, separation, and the restriction of all branches or a specific branch to the real or imaginary axis. Here, we use the biorthogonal framework to reveal the origin of these structures and determine the corresponding critical momentum conditions by analyzing the values of $m_k=\hat{\mathbf{h}}_0\cdot\hat{\mathbf{h}}_1$, the dot product between the normalized prequench and postquench complex Bloch vectors. When Fisher zeros continuously cover the real-time axis, the density of Fisher zeros is further introduced to identify the critical times within these critical regions, at which the rate function becomes nonanalytic. We verify our results in a generalized non-Hermitian SSH model by quenching the Peierls phase and the non-Hermitian gain-loss strength, and also use the Jordan form to analyze the dynamics at postquench exceptional points, where the conventional Fisher zero formula is not applicable.

\end{abstract}

\maketitle

\section{Introduction}
Dynamical quantum phase transitions (DQPTs) are critical phenomena emerging in the time domain during nonequilibrium quantum evolution, characterized by the nonanalytic behavior of the rate function corresponding to the Loschmidt echo in the thermodynamic limit\cite{PhysRevLett.110.135704,heyl2018dynamical}. This concept was first proposed in the quench dynamics of the transverse-field Ising model, and subsequently used to study dynamical critical behavior, dynamical topological properties, and connections between DQPTs and equilibrium phase transitions\cite{PhysRevLett.110.135704,heyl2018dynamical,budich2016dynamical,Zvyagin,Heyl_2019,bhattacharya2017emergent,vajna2014disentangling,vajna2015topological,wong2024entanglement,Chenshu}. The study of DQPTs has also been extended beyond sudden quenches in closed systems to include slow quenches\cite{sharma2016slow,PhysRevB.95.184307,zamani2024scaling,baghran2024competition,PhysRevB.109.L180303,cao2024exploring,32c2-z9hf}, phase quenches \cite{PhysRevB.106.045410,nie2024flux,zhang2025phase,nie2026Floquet}, Floquet systems\cite{PhysRevB.100.085308,PhysRevB.102.144306,zhou2021floquet,PhysRevA.103.012204,luan2022floquet,PhysRevB.106.094314,PhysRevB.105.094311}, mixed states\cite{PhysRevB.96.180303,PhysRevB.96.180304,cao2025}, and open or dissipative systems\cite{PhysRevB.98.134310,PhysRevA.105.022220,PhysRevB.97.045147,gqv9-rfss,xu2026}. Meanwhile, experimental studies have been performed in trapped ions, cold atomic gases, and quantum simulation platforms\cite{PhysRevLett.119.080501,PhysRevB.100.085308,Flaeschner2018,guo2019observation,PhysRevLett.122.020501,Xu2020}.

In recent years, research on DQPTs has been extended to non-Hermitian systems\cite{PhysRevA.98.022129,PhysRevLett.122.020501,Zhou_2021,PhysRevB.106.054308,PhysRevB.107.184311,Mondal2024,PhysRevLett.132.220402,92w1-71rt,Zhang2025SelfNormal,f5k4-l93p}. Since non-Hermitian evolution is generally nonunitary and the left and right eigenstates are generally different, the definition of the Loschmidt echo and its probability interpretation have become important issues.
Early studies employed the self-normal framework, which uses the conventional inner product and normalizes the time-evolved right state to characterize its probability of returning to the initial state\cite{PhysRevA.98.022129,PhysRevLett.122.020501,Zhou_2021,PhysRevB.106.054308,PhysRevB.107.184311,Mondal2024}. Subsequently, a framework for biorthogonal DQPTs was proposed based on biorthogonal bases and associated states\cite{PhysRevLett.132.220402}, in which the Loschmidt echo is defined as a real probability ranging from 0 to 1 and applies to non-Hermitian systems with complex eigenvalues. The two frameworks use different inner products and may therefore yield different critical momenta and critical times. Recent experiments on non-Hermitian quantum walks with single photons have compared the rate functions and dynamical topological order parameters in the two frameworks and observed different DQPT critical times\cite{Zhang2025SelfNormal}. To ensure that the description of quench dynamics is consistent with the left and right eigenstate structure of non-Hermitian systems and the corresponding probability definition, we adopt the biorthogonal framework in this work.

The analysis of Fisher zeros provides an important approach to determining the critical conditions for DQPTs and understanding the nonanalyticity of the rate function\cite{PhysRevLett.110.135704,heyl2018dynamical,fisher1965nature,PhysRevB.92.075114,W1984,PhysRevB.110.224302}. By analogy with the analysis of zeros of the complex partition function in equilibrium phase transitions, DQPTs are studied by extending the time variable $t$ to the complex time plane and finding the Fisher zeros of the Loschmidt amplitude. In the thermodynamic limit, DQPTs occur when the lines or areas formed by these zeros hit the real-time axis. For a Hermitian two-band system, this condition requires the initial lower Bloch state at a critical momentum $k_c$ to be an equal weight superposition of the two postquench Bloch eigenstates, which is equivalent to $m_{k_c}=\hat{\mathbf{h}}_{k_c}^{(0)}\cdot \hat{\mathbf{h}}_{k_c}^{(1)}=0$, i.e., the dot product between the normalized prequench and postquench Bloch vectors vanishes, and then the critical time $\tau_c$ is determined by the spectrum of the postquench Hamiltonian at the critical momentum $k_c$\cite{vajna2015topological,guo2019observation,PhysRevB.97.060304,Chenshu,nie2024flux}. However, in non-Hermitian systems, $m_k$ and the postquench spectrum may each appear in both the critical momentum and critical time conditions\cite{PhysRevLett.132.220402,92w1-71rt}. Since the Bloch vectors are generally complex, $m_k$ is no longer restricted to $[-1,1]$ and may take real values ($|m_k|>1$), purely imaginary, or generic complex values. We will show that the unique properties of $m_k$ in non-Hermitian systems lead to anomalous structures of Fisher zero lines, including divergence, separation, and the restriction of all branches or a specific branch to the real or imaginary axis of the complex-time plane, which present new challenges for identifying non-Hermitian DQPTs. When Fisher zeros continuously cover the imaginary axis, the existence of zeros is insufficient to determine the DQPT critical times, and it is necessary to analyze their distribution and its relation to the nonanalyticity of the rate function. Therefore, we will introduce the density of Fisher zeros to connect the analysis of anomalous structures of Fisher zero lines with the determination of DQPT critical times. 

The rest of this paper is arranged as follows. In Sec.~\ref{sec2}, we will introduce the biorthogonal Loschmidt echo and the general expression for Fisher zeros in two-band non-Hermitian systems, and then analyze the critical momentum conditions for DQPTs as well as the anomalous structures of Fisher zero lines by considering the different values of $m_k=\hat{\mathbf{h}}_0\cdot\hat{\mathbf{h}}_1$ together with the energy spectrum after quenching. The density of Fisher zeros will be introduced to distinguish critical regions continuously covered by Fisher zeros from the critical times, at which the rate function becomes nonanalytic. In Sec.~\ref{sec3}, we will verify our results by considering the DQPTs emerging in the quenched generalized non-Hermitian Su-Schrieffer-Heeger(SSH) model with a Peierls phase. The case in which the postquench Hamiltonian reaches EPs at isolated momenta will also be considered and various anomalous behaviors of the Fisher zero lines will be discussed. Finally, we summarize our results in Sec.~\ref{sec4}.

\section{Fisher Zeros in the Biorthogonal Framework}\label{sec2}
To study the behavior of quenched non-Hermitian two-band systems, we first introduce the biorthogonal framework and the biorthogonal Loschmidt echo. 

For a general system with non-Hermitian Hamiltonian $H$, the left and right eigenstates are defined as\cite{Brody_2014}
\begin{equation}
	H|u_{n}\rangle=E_{n}|u_{n}\rangle , \qquad
    H^{\dagger}| \tilde{u}_n \rangle      =E_{n}^{*}| \tilde{u}_n \rangle,    
\end{equation}
where $E_{n}$ is the $n$-th eigenvalue. When the Hamiltonian is away from EPs, the left eigenstates $|\tilde{u}_n \rangle $ and the right  eigenstates $|u_{n}\rangle$ satisfy the biorthogonal relation $\langle\widetilde{u}_{m}|u_{n}\rangle=\delta_{mn}$ and the completeness relation $\sum_{n}|u_{n}\rangle\langle\widetilde{u}_{n}|=I$.  

Consider a sudden quench from  $H_k^{(0)}$ to $H_k^{(1)}$ in a translationally invariant non-Hermitian two-band lattice system, where the prequench and postquench Bloch Hamiltonians are given by $H_k^{(s)}=\mathbf{h}_k^{(s)}\cdot\boldsymbol{\sigma}$, with $s=0, 1$, $\boldsymbol{\sigma}=(\sigma_x,\sigma_y,\sigma_z)$ being the vector of Pauli matrices and $\mathbf{h}_k^{(s)}$ a complex Bloch vector.  
The normalized complex Bloch vector is defined as 
$\hat{\mathbf{h}}_k^{(s)}=\mathbf{h}_k^{(s)}/h_k^{(s)}$ with $h_k^{(s)}
=\sqrt{(h_{k,x}^{(s)})^2+(h_{k,y}^{(s)})^2+(h_{k,z}^{(s)})^2}$, and 
the  eigenvalues of $H^{(s)}_k$ and $H^{\dagger(s)}_k$ are $E_{k,\pm}^{(s)}=\pm h^{(s)}_k$ and $E_{k,\pm}^{*(s)}=\pm h^{*(s)}_k$. 
For an arbitrary given state $|\psi \rangle =\sum_{\pm}{c_{\pm}}|u_{\pm} \rangle $ with $c_{\pm}=\langle \tilde{u}_{\pm}|\psi \rangle$, and its associated state $| \tilde{\psi}\rangle:=\sum_{\pm}{c_{\pm}}|\tilde{u}_{\pm}\rangle$ \cite{Brody_2014}, the biorthogonal Loschmidt echo between the initial state $| \psi (0) \rangle $ and the time-evolved state $| \psi (t) \rangle $ is defined as\cite{PhysRevLett.132.220402}
\begin{equation}\label{BLE}
\mathcal{L}(t) =\frac{\langle \tilde{\psi}(0) |\psi(t)\rangle \langle \tilde{\psi}(t) |\psi (0)\rangle }{\langle \tilde{\psi}(t) |\psi(t)\rangle \langle \tilde{\psi}(0) |\psi(0)\rangle}.
\end{equation}  
In this work, the prequench Hamiltonian is chosen away from EPs for all momenta to ensure that the initial biorthogonal eigenbasis is well defined. Assuming that the initial state is $|\Psi(0)\rangle=\bigotimes_k|u_{k,-}^{(0)}\rangle$, the corresponding Loschmidt echo can be obtained as 
\begin{eqnarray}\label{LE}
\mathcal{L}\left( t \right) &=&\prod_k{\mathcal{L}_k}\left( t \right) =\prod_k{\frac{|\langle       \tilde{u}_{k,-}^{(0)}|e^{-i H_k^{(1)} t}|u_{k,-}^{(0)} \rangle|^2}{\langle       \tilde{u}_{k,-}^{(0)}(t)|u_{k,-}^{(0)}(t) \rangle \langle\tilde{u}_{k,-}^{(0)}| u_{k,-}^{(0)} \rangle }}      \nonumber  \\
&=&\prod_k{\frac{|b_k(t)|^2}{|a_k(t)|^2+|b_k(t)|^2}}, 
\end{eqnarray}
where $a_k(t)=\langle \tilde{u}_{k,+}^{(0)}|e^{-i H_k^{(1)}t}|u_{k,-}^{(0)} \rangle $
and the Loschmidt amplitude for each momentum mode $k$ is given by
\begin{eqnarray}
G_k(t) &=& b_k(t) =\langle       \tilde{u}_{k,-}^{(0)}|e^{-i H_k^{(1)} t}|u_{k,-}^{(0)} \rangle     \nonumber \\  
&=&\cos ( h^{(1)}_k   t) +i\sin (h^{(1)}_k   t) m_k, 
\end{eqnarray}
where $m_k=\mathbf{\hat{h}}_{k}^{(0)}\cdot \mathbf{\hat{h}}_{k}^{(1)}=( \mathbf{h}_{k}^{(0)}\cdot \mathbf{h}_{k}^{(1)} ) /( h_{k}^{(0)}h_{k}^{\left( 1 \right)} )$ is generally complex value in non-Hermitian systems.

\subsection{Anomalous structure of Fisher zero lines}\label{Bpart}
In order to investigate the zero structure of the Loschmidt amplitude, we extend the time variable to the complex plane with $z=\tau+it\in\mathbb{C}$, and the Fisher zero lines can be written as\cite{92w1-71rt}
\begin{equation}\label{GFZ}
z_n\left( k \right) =\frac{ia_n-Q_k}{h^{(1)}_k}, 
\end{equation}
where $Q_k=\tanh^{-1}m_k$, $a_n=(n+1/2)\pi$, $n\in \mathbb{Z}$. DQPTs occur when the Fisher zero lines cross the real-time axis, i.e., when $\operatorname{Re}z_n(k_c)=0$, with the corresponding critical time $t_c=\operatorname{Im}z_n(k_c)$. Note that Eq.(\ref{GFZ}) is applicable to momentum modes when $h_k^{(1)}\neq 0$ and the postquench Hamiltonian $H_k^{(1)}$ is diagonalizable. When $H_k^{(1)}$ reaches an EP at a given momentum, $h_k^{(1)}=0$, so $m_k$ and $Q_k$ are not defined and Eq.~(\ref{GFZ}) cannot be applied directly. The postquench dynamics at EPs will be discussed using the Jordan form in Sec.~\ref{3.3}.

The above expression indicates that the expressions for Fisher zeros in non-Hermitian systems within the biorthogonal framework contain the dot product $m_k=\mathbf{\hat{h}}_{k}^{(0)}\cdot \mathbf{\hat{h}}_{k}^{(1)}$ between the normalized prequench and postquench Bloch vectors.
Since the Bloch vectors are generally complex in non-Hermitian systems, $m_k$ is no longer constrained to $[-1,1]$ and can take real values ($|m_k|>1$), purely imaginary values, or generic complex values. The unique properties of $m_k$ and the postquench energy spectrum $h_k^{(1)}$ in non-Hermitian systems lead to anomalous Fisher zero structures, which are discussed below.

(i) Generic complex $m_k$ with nonzero real and imaginary parts and $Q_k=Q_{k,\mathrm{R}}+iQ_{k,\mathrm{I}}$. 
If the postquench spectrum is complex, i.e., $h_k^{(1)}=h_{k,\mathrm{R}}^{(1)}+i h_{k,\mathrm{I}}^{(1)}$ ($h_{k,\mathrm{R}}^{(1)},h_{k,\mathrm{I}}^{(1)} \in \mathbb{R}$), Eq.~(\ref{GFZ}) can be expressed as
\begin{eqnarray}
z_n=\frac{h_{k,\mathrm{I}}^{(1)}(a_n-Q_{k,\mathrm{I}})-h_{k,\mathrm{R}}^{(1)}Q_{k,\mathrm{R}}}{|h_k^{(1)}|^2} \\ \nonumber
\ \ \ \ \   +i\frac{h_{k,\mathrm{R}}^{(1)}(a_n-Q_{k,\mathrm{I}})+h_{k,\mathrm{I}}^{(1)}Q_{k,\mathrm{R}}}{|h_k^{(1)}|^2}.
\end{eqnarray}
The occurrence of DQPT requires $\operatorname{Re}z_n=0$, and the critical momentum condition reduces to $h_{k,\mathrm{I}}^{(1)}(a_n-Q_{k,\mathrm{I}})=h_{k,\mathrm{R}}^{(1)}Q_{k,\mathrm{R}}$, consistent with Ref.~\cite{92w1-71rt}, in which the Fisher zero trajectories are distributed over the complex time plane. However, we find that richer structures appear when $m_k$ is real or purely imaginary.

(ii)  Real number of $m_k \in \mathbb{R}$, with the form of $Q_k$ depending on $|m_k|$.

(a) If $|m_k|<1$, then $Q_k=Q_{k,\mathrm{R}} \in \mathbb{R}$ and $Q_{k,\mathrm{I}}=0$. The critical momentum condition reduces to $h_{k,\mathrm{I}}^{(1)}a_n=h_{k,\mathrm{R}}^{(1)}Q_{k,\mathrm{R}}$. If the postquench spectrum is also real, $h_k^{(1)}=h_{k,\mathrm{R}}^{(1)}$, the critical momentum condition is reduced to $m_{k_c}=0$ and the corresponding critical time is $t_{c,n}=a_n/h_{k_c,\mathrm{R}}^{(1)}$, which reduces to the critical condition of conventional DQPT in the Hermitian limit. For a purely imaginary postquench spectrum, $h_k^{(1)}=ih_{k,\mathrm{I}}^{(1)}$, $\operatorname{Re}z_n={a_n}/{h_{k,\mathrm{I}}^{(1)}}\ne 0$. 
In this case, the Fisher zero line does not cross the imaginary axis, and no corresponding finite critical time exists. 

(b) If $m_k\rightarrow\pm1$ along the real axis from within $(-1,1)$, then $Q_{k,\mathrm{I}}=0$ and $|Q_{k,\mathrm{R}}|\rightarrow\infty$.  For a generic complex postquench spectrum $h_k^{(1)}=h_{k,\mathrm{R}}^{(1)}+ih_{k,\mathrm{I}}^{(1)}$, the Fisher zeros satisfy  $\operatorname{Re}z_n\approx -{h_{k,\mathrm{R}}^{(1)}Q_{k,\mathrm{R}}}/{|h_k^{(1)}|^2}$ and $\operatorname{Im}z_n\approx{h_{k,\mathrm{I}}^{(1)}Q_{k,\mathrm{R}}}/{|h_k^{(1)}|^2}$. Therefore, both the real and imaginary parts of the Fisher zeros generally diverge, with the asymptotic direction determined by the $\operatorname{Im}z_n/\operatorname{Re}z_n\rightarrow -h_{k,\mathrm{I}}^{(1)}/h_{k,\mathrm{R}}^{(1)}$. Hence, the Fisher zero line extends to infinity along a direction determined by the phase of the complex postquench spectrum. For a purely real spectrum, the divergence occurs along the real direction, $|\operatorname{Re}z_n|\rightarrow\infty$, while $\operatorname{Im}z_n$ remains finite. For a purely imaginary spectrum, $|\operatorname{Im}z_n|\rightarrow\infty$, whereas $\operatorname{Re}z_n=a_n/h_{k,\mathrm{I}}^{(1)}$ remains finite but nonzero. Therefore, these momentum modes do not lead to a DQPT at any finite time in the limit $m_k\to\pm1$. As $m_k\rightarrow \pm 1$ approaches $\pm1$ from the region $|m_k|>1$ , $Q_{k,\mathrm{I}}$ is nonzero but remains finite. Therefore, the divergence of the Fisher zero lines is still controlled by $Q_{k,\mathrm{R}}$.

(c) If $|m_k|>1$, both $Q_{k,\mathrm{R}}$ and $Q_{k,\mathrm{I}}$ are nonzero, such that
\begin{equation}
Q_k=Q_{k,\mathrm{R}}+iQ_{k,\mathrm{I}}=\frac{1}{2}\ln\text{|}\frac{1+m_k}{1-m_k}|+i\frac{\left( 2l+1 \right) \pi}{2}  \ \ l\in \mathbb{Z}.
\end{equation}
For a generic complex postquench spectrum with $h_k^{(1)}=h_{k,\mathrm{R}}^{(1)}+i h_{k,\mathrm{I}}^{(1)}$, the critical momentum condition is reduced to $\left( n-l \right) \pi h_{k,\mathrm{I}}^{(1)}=h_{k,\mathrm{R}}^{(1)}Q_{k,\mathrm{R}}$, in which $l$ labels the logarithmic branch of $Q_k$, whereas $n$ labels the Fisher zero branches. For a purely real postquench spectrum $h_k^{(1)}=h_{k,\mathrm{R}}^{(1)}$, the mode does not satisfy  $Q_{k,\mathrm{R}}=0$ if $|m_k|>1$ and does not generate a finite time DQPT. In particular, for $n=l$, we have $\operatorname{Im}z_n=0$, and corresponding Fisher zero branch is confined to the real axis of the complex time plane. For a purely imaginary postquench spectrum $h_k^{(1)}=ih_{k,\mathrm{I}}^{(1)}$, the condition $\operatorname{Re}z_n=0$ requires $n=l$. The corresponding Fisher zero branch is confined to the imaginary axis of the complex time plane.

(iii) Purely imaginary $m_k=i\alpha _k$ with $\alpha _k\in \mathbb{R}$ and $\alpha_k\neq0$ and $Q_k=iQ_{k,\mathrm{I}}=i\left( \arctan \alpha _k +l\pi \right)$ with $l\in \mathbb{Z}$. Since $\operatorname{Re}z_n=0$ for a purely real postquench spectrum and $\operatorname{Im}z_n=0$ for a purely imaginary one, all Fisher zero branches are restricted to the imaginary axis and real axis of the complex time plane, respectively.

The above categories show that the values of $m_k$ remain essential for understanding the anomalous structures of Fisher zero lines, although $m_{k_c}=\mathbf{\hat{h}}_{k_c}^{(0)}\cdot \mathbf{\hat{h}}_{k_c}^{(1)}=0$ is no longer a general critical momentum condition in non-Hermitian systems. The different values and limits of $m_k$, together with the postquench spectrum, explain why Fisher zero lines diverge, separate, or become restricted to the real or imaginary axis.

\subsection{Density of Fisher zeros and nonanalyticities in the rate function}
For the typical one-dimensional Hermitian case, the Fisher zero lines usually cross the real-time axis at isolated critical momenta, giving discrete critical times at which DQPTs occur. However, for the quenched dynamics of non-Hermitian system, either all branches of the Fisher zero line or a specific branch lie on the real-time axis, forming a critical times region similar to that in two-dimensional Hermitian systems\cite{PhysRevB.109.134301,PhysRevB.110.224302}. Within these regions, the time corresponding to each momentum mode satisfies the critical condition, but not every time within these region corresponds to a critical time at which the DQPT occurs. Therefore, we introduce the the density of Fisher zeros to characterize the distribution of these zeros along the real-time axis and to identify the critical times at which the rate function becomes nonanalytic.

By using the biorthogonal Loschmidt amplitude and Loschmidt echo defined above, the rate function can be expressed as
\begin{equation}
r(t) =-\underset{N\rightarrow \infty}{\lim}\frac{1}{N}\ln \mathcal{L}\left( t \right) =2\operatorname{Re}f\left( t \right) +\underset{N\rightarrow \infty}{\lim}\frac{1}{N}\ln \mathcal N(t)
\end{equation}
where $f(t) =-\underset{N\rightarrow \infty}{\lim}\ln G(t)/N$ and $\mathcal N(t)=\prod_k\mathcal N_k(t)$ is the normalization factor in the biorthogonal Loschmidt echo. Since each mode-resolved normalization factor $\mathcal N_k(t)$ is positive and smooth when the initial Hamiltonian is away from EPs, the nonanalyticities in the rate function are determined solely by the zeros of the Loschmidt amplitude. Consider the Loschmidt amplitude with time extended into the complex plane. For a finite system, $G(z)$ is an entire function of the complex variable $z$ and its zeros $z_j$ can be explicitly separated using the Weierstrass factorization\cite{PhysRevLett.110.135704}
\begin{equation}
G\left( z \right) =e^{g\left( z \right)}\prod_j{\left( 1-\frac{z}{z_j} \right)}.
\end{equation}
With this, the dynamical free energy density can be obtained,
\begin{equation}
f\left( z \right) =-\underset{N\rightarrow \infty}{\lim}\frac{1}{N}\left[ g\left( z \right) +\sum_j{\ln \left( 1-\frac{z}{z_j} \right)} \right]. 
\end{equation}
and its nonanalytic behavior is contained in the logarithmic term. In the thermodynamic limit, the sum over discrete zeros can be transformed into an integral with respect to the continuous variable $x\in X$. Denoting the corresponding position of a Fisher zero in the complex time plane by $\tilde z(x)$, the singular part of the dynamical free energy density can be written as\cite{PhysRevB.92.075114}
\begin{equation}
\tilde{f}(z) {=}{-}\int_X{dx}\ln ( 1{-}\frac{z}{\tilde{z}(x)} ) {=}{-}\int_{\tilde{z}(X)}{d\tilde{z}\rho ( \tilde{z})}\ln (1{-}\frac{z}{\tilde{z}}),
\end{equation}
where the second equality follows from a transformation of the integration variable, and $\rho(\tilde z)$ is determined by the corresponding Jacobian which represents the density of Fisher zeros in the complex time plane \cite{W1984,PhysRevB.92.075114,PhysRevB.110.224302}. By taking the real part, the zero-dependent contribution to the rate function becomes
\begin{equation}
\tilde{r}\left( z \right) =2\operatorname{Re}\left[ \tilde{f}\left( z \right) \right] =-2\int_{\mathbb{C}}{d\tilde{z}\rho \left( \tilde{z} \right)}\ln |1-\frac{z}{\tilde{z}}|.
\end{equation}
According to Ref. \cite{PhysRevB.92.075114}, $\operatorname{Re}\tilde f(z)$ can be regarded as a two-dimensional electrostatic potential generated by the density of Fisher zeros $\rho(z)$, which acts as an effective charge density. 
This analogy suggests that the nonanalytic behavior of the rate function is directly related to the nonanalytic structures of the density of Fisher zeros\cite{heyl2018dynamical}.
For the one-dimensional cases considered here, the relevant Fisher zeros lie on the real-time axis, satisfying $\operatorname{Re}z_n(k)=0$, $t_n(k)=\operatorname{Im}z_n(k)$. 
In this case, the line density of Fisher zeros $\rho (t)$ along this axis can be defined as
\begin{equation}
\rho \left( t \right) =\sum_n{\int{\frac{dk}{2\pi}\delta \left[ t-\operatorname{Im}z_n\left( k \right) \right]}}, 
\end{equation}
which represents the contribution of all Fisher zero branches along the real-time axis\cite{PhysRevB.110.224302}. 

Therefore, the density of Fisher zeros not only describes their distribution within the critical regions but is also linked to the nonanalyticities in the rate function, thereby allowing the distinction between the critical region where the Fisher zeros lie on the real-time axis and the critical times at which the DQPT occurs.

\section{Anomalous Fisher zero structures in a quenched generalized non-Hermitian SSH model with a Peierls phase}\label{sec3}
Now, we will consider a generalized non-Hermitian SSH model with a Peierls phase $\phi$, which can be modulated by a synthetic gauge field\cite{Fang2012,Schmidt15,PhysRevA.100.012112}, and show that quenching $\phi$ and the balanced gain-loss strength $\gamma$ can induce the anomalous structures of Fisher zero lines when $m_k$ is real or purely imaginary.

\subsection{Model and PT phase diagram}
\begin{figure}
	\centering
	{\includegraphics[width=1.1\columnwidth]{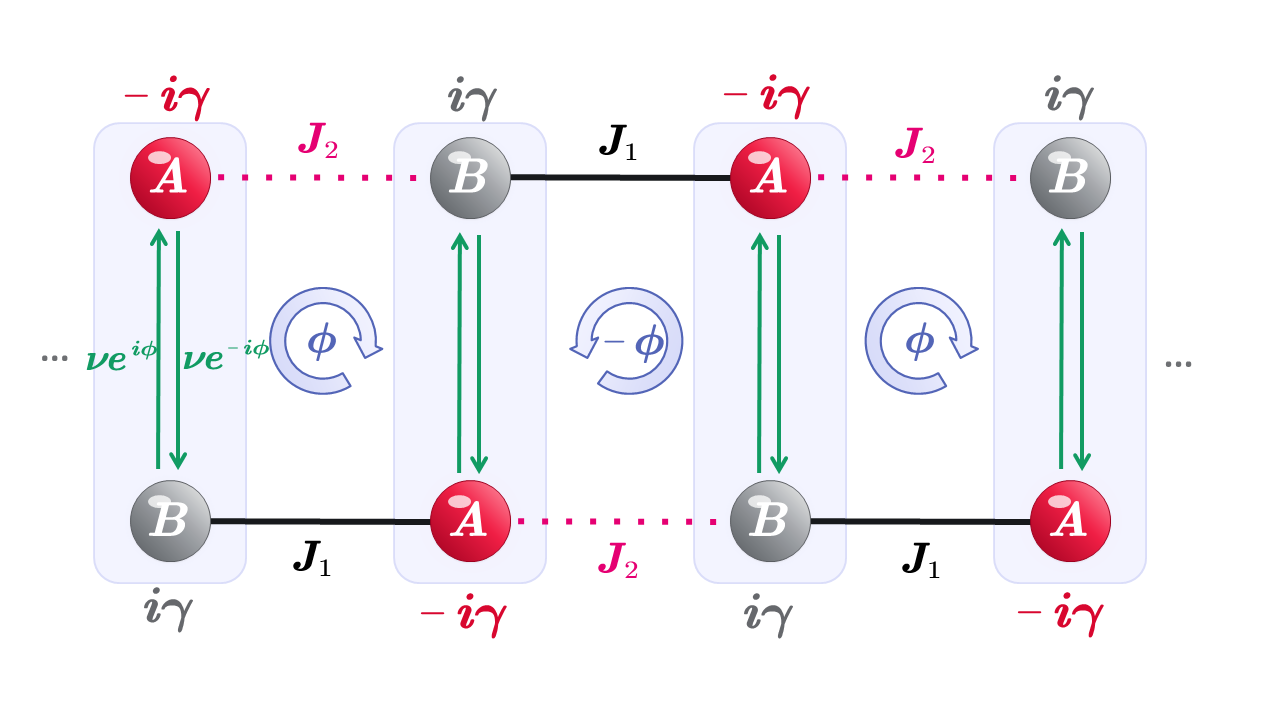}}
	\caption{The ladder lattice of the generalized non-Hermitian SSH model with a Peierls phase and gain and loss. The green arrows represent the intracell hoppings $ve^{\pm i\phi}$, while the black solid and pink dashed lines denote the real intercell couplings $J_1$ and $J_2$, respectively. The blue circular arrows represent synthetic gauge fields with opposite directions in adjacent plaquettes, allowing the Peierls phase $\phi$ to be controlled. The on-site potentials $-i\gamma$ and $i\gamma$ on sites A and B describe balanced loss and gain.
	} 
	\label{Fig1}
\end{figure}
As illustrated in Fig.~\ref{Fig1}, the generalized non-Hermitian Su-Schrieffer-Heeger (SSH) model with a tunable Peierls phase contains two real intercell couplings $J_1$ and $J_2$, complex intracell coupling $ve^{i\phi}$, and on-site potentials on sublattices $A$ and $B$, where $\gamma$ denotes the strength of the balanced loss and gain\cite{PhysRevA.100.012112}. The uniform Peierls phase $\phi$ can be tuned via a synthetic gauge field whose direction alternates between adjacent plaquettes.
\begin{figure}
	\centering
	{\includegraphics[width=1.1\columnwidth]{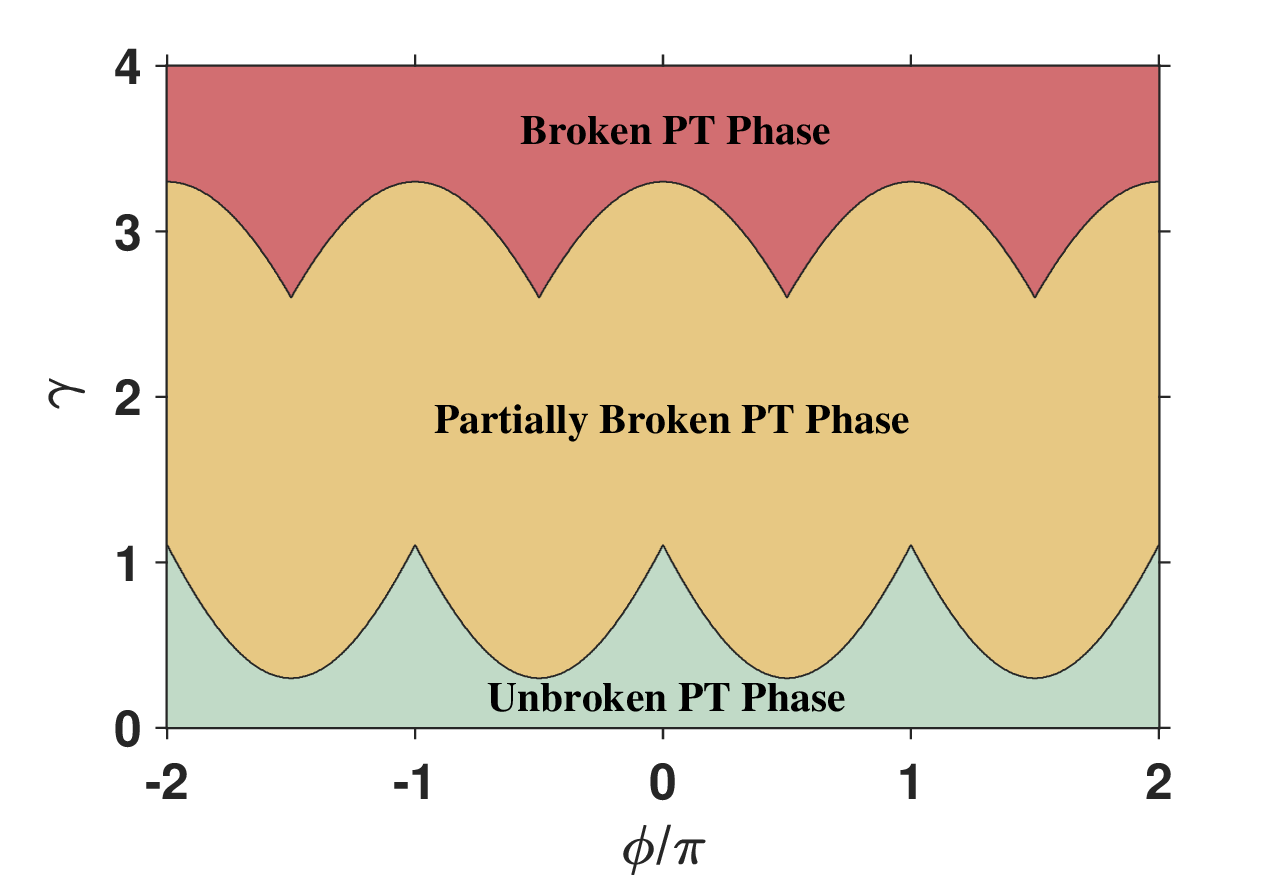}}
	\caption{PT phase diagram in the $(\phi,\gamma)$ plane for $v=1$, $J_1=1.8$, and $J_2=0.5$. The green, yellow, and red regions denote the unbroken, partially broken, and broken PT phases, respectively.
	} 
	\label{Fig2}
\end{figure}
As discussed in Ref.~\cite{PhysRevA.100.012112}, this model does not exhibit the non-Hermitian skin effect, so its bulk properties can be described directly using the Bloch Hamiltonian.
For a translationally invariant system, the Bloch Hamiltonian in $k$-mode subspace can be expressed as 
$H_k(\phi ) =\boldsymbol{h}_{k}(\phi) \cdot \boldsymbol{\sigma }$
with
\begin{eqnarray}
&&h_{k,x}(\phi) = v\cos \phi +\left( J_1+J_2 \right) \cos k,  \nonumber \\
&&h_{k,y}(\phi ) = -v\sin \phi +\left( J_1-J_2 \right) \sin k,  \nonumber \\
&&h_{k,z}(\phi)=-i\gamma .
\end{eqnarray}
The eigenvalues of $H_k(\phi )$ are given by
\begin{equation}
E_{k,\pm}^{}=\pm h_k(\phi) =\pm \sqrt{h_{k,x}^2+h_{k,y}^2-\gamma ^2}.
\end{equation}
and the spectrum of each momentum mode is either purely real or purely imaginary, except at the EPs where $h_k=0$. For $\gamma\neq0$, the model is non-Hermitian and respects
parity-time (PT) symmetry. Fig.~\ref{Fig2} shows the corresponding PT phase diagram in the $(\phi,\gamma)$ plane for $v=1$, $J_1=1.8$, and $J_2=0.5$, where a nonzero $J_2$ is chosen so that the energy bands have a nontrivial dependence on the Peierls phase $\phi$, and avoid the special case $J_1=J_2$, for which the $k$ dependence of $h_y$ vanishes\cite{PhysRevA.100.012112}. The green region denotes the unbroken PT phase, where the spectra are real for all momenta. The yellow region corresponds to the partially broken PT phase, where the spectra are real for some momenta and purely imaginary for others. The red region represents the fully broken PT phase, where the spectra are purely imaginary for all momenta. The model also supports topologically distinct phases characterized by a winding number, and the Peierls phase $\phi$ can change the winding number, whereas the gain-loss strength $\gamma$ controls the PT symmetry transition without changing the topological phase boundary.

\subsection{Quench dynamics away from exceptional points}

Now we consider a sudden quench of the Peierls phase and the non-Hermitian gain-loss strength, $(\phi_i,\gamma_i)\rightarrow (\phi_f,\gamma_f)$, and denote the prequench and postquench Bloch vectors by $ \mathbf h^s_k = (h^s_{k,x},h^s_{k,y},-i\gamma ) $. For this model, the overlap $\mathbf{h}_{k}^{\left( 0 \right)}\cdot \mathbf{h}_{k}^{\left( 1 \right)}=h_{k,x}^{\left( 0 \right)}h_{k,x}^{\left( 1 \right)}+h_{k,y}^{\left( 0 \right)}h_{k,y}^{\left( 1 \right)}-\gamma ^{\left( 0 \right)}\gamma ^{\left( 1 \right)}\in \mathbb{R}$, hence $m_k$ is real if the prequench and postquench spectra are both real or both purely imaginary, whereas $m_k$ is purely imaginary if one spectrum is real and the other is purely imaginary. Accordingly, we first consider the unbroken-to-unbroken (UU), unbroken-to-broken (UB), broken-to-unbroken (BU), and broken-to-broken (BB) cases, depending on whether the prequench and postquench Hamiltonians are in the unbroken PT phase or fully broken PT phase. The unbroken-to-partially-broken (UP) and broken-to-partially-broken (BP) cases, including the dynamics at the EPs of the postquench Hamiltonian in the partially broken PT phase, will be discussed in the next subsection.

\begin{figure}
	\centering
	{\includegraphics[width=1\columnwidth]{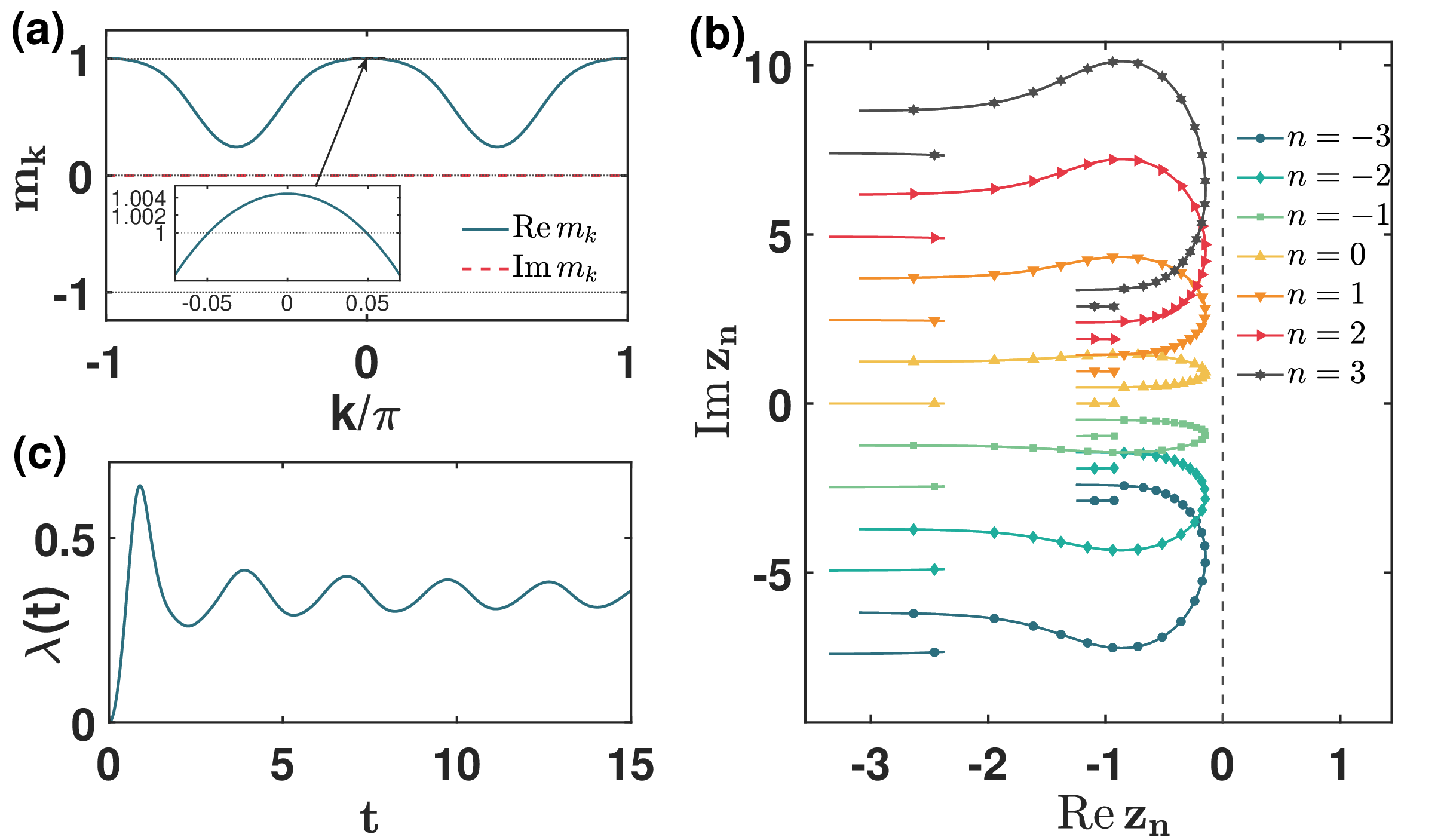}}
	\caption{Results for the UU quench from $(\phi_i,\gamma_i)=(0,0.2)$ to $(\phi_f,\gamma_f)=(\pi,0.2)$. (a) The real and imaginary parts of $m_k$ as functions of $k$, where the blue solid and red dashed lines denote $\operatorname{Re}m_k$ and $\operatorname{Im}m_k$, respectively. (b) Fisher zeros $z_n(k)$ in the complex time plane for different $n$, with different colors and symbols representing different branches. The vertical dashed line marks $\operatorname{Re}z=0$. (c) The rate function $\lambda(t)$ as a function of time $t$.
	} 
	\label{Fig3}
\end{figure}
For the UU quench, the Peierls phase is quenched from $\phi_i=0$ to $\phi_f=\pi$ while keeping $\gamma_i=\gamma_f=0.2$ fixed. In this case, both the initial and final spectra are purely real for all momenta, i.e., $h^{(0)}_k$, $h^{(1)}_k$, and $m_k$ are real. As shown in Fig.~\ref{Fig3}(a), $\operatorname{Im}m_k=0$, while $\operatorname{Re}m_k$ remains positive throughout the Brillouin zone and does not cross zero, and regions with $m_k>1$ are present near $k=0$ and the Brillouin-zone boundaries. As discussed above in Sec.~\ref{Bpart}, real $m_k$ in non-Hermitian systems can give rise to anomalous behavior of the Fisher zero lines. When $m_k\rightarrow 1$, $|Q_{k,\mathrm{R}}|\rightarrow\infty$. For the UU quench, $h^{(1)}_k$ is real, so that $\operatorname{Re}z_n\rightarrow -\infty$. Consequently, as shown in Fig.~\ref{Fig3}(b), the branches of the Fisher zeros extend toward $\operatorname{Re}z\rightarrow-\infty$ as $m_k\rightarrow 1$ and return to the complex time region when $m_k>1$. Since only a limited region of the complex time plane can be displayed, these lines appear to be disconnected in Fig.~\ref{Fig3}(b). When $m_k>1$, $Q_{k,\mathrm{I}}$ additionally acquires a branch-dependent imaginary part, which modifies the imaginary coordinates of the corresponding Fisher zeros and gives rise to the separated branch structure observed in Fig.~\ref{Fig3}(b), with part of the $n=0$ Fisher zero branch lying on the real axis for the principal branch $l=0$. However, all Fisher zeros remain on the $\operatorname{Re}z<0$ side and do not reach the imaginary axis. Consistently, the rate function shown in Fig.~\ref{Fig3}(c) evolves smoothly without any nonanalytic cusp. Therefore,  despite the UU quench exhibits the characteristic anomalous structure of the Fisher zeros associated with real $m_k$ in the non-Hermitian system, no DQPT occurs for the parameters considered here. 

\begin{figure}
	\centering
	{\includegraphics[width=1\columnwidth]{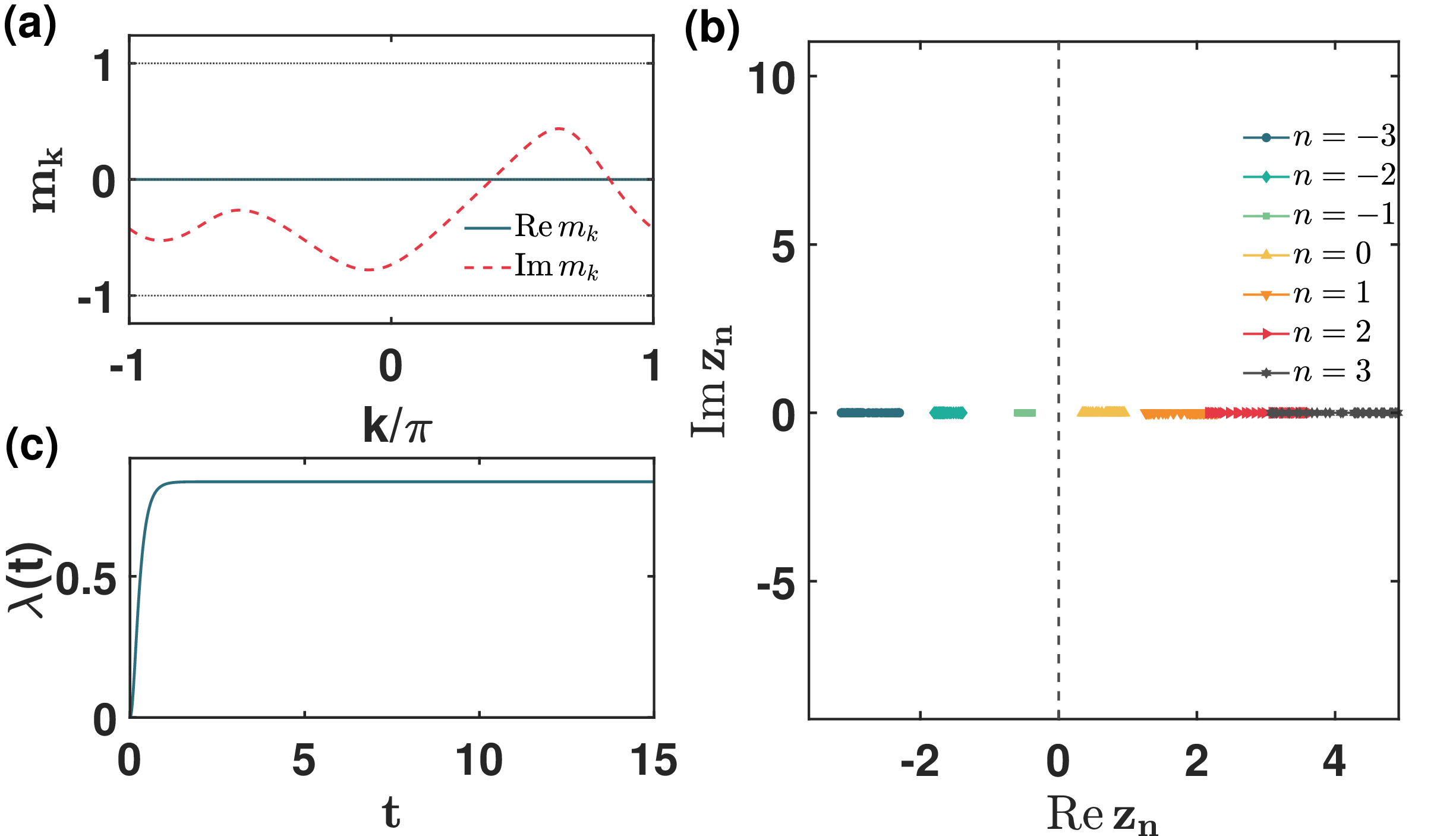}}
	\caption{Results for the UB quench from $(\phi_i,\gamma_i)=(0,0.5)$ to $(\phi_f,\gamma_f)=(0.5\pi,3.5)$. (a) The real and imaginary parts of $m_k$ as functions of $k$, where the blue solid and red dashed lines denote $\operatorname{Re}m_k$ and $\operatorname{Im}m_k$, respectively. (b) Fisher zeros $z_n(k)$ in the complex time plane for different $n$, with different colors and symbols representing different branches. The vertical dashed line marks $\operatorname{Re}z=0$. (c) The rate function $\lambda(t)$ as a function of time $t$.
	} 
	\label{Fig4}
\end{figure}

For the UB quench from $(\phi_i,\gamma_i)=(0,0.5)$ to $(\phi_f,\gamma_f)=(0.5\pi,3.5)$, the initial spectrum is purely real, whereas the final spectrum is purely imaginary, so $m_k$ is purely imaginary, i.e., $\operatorname{Re}m_k=0$ throughout the Brillouin zone, as shown in Fig.~\ref{Fig4}(a). According to the discussion above for purely imaginary $m_k$, $Q_k=\tanh^{-1}m_k$ is also purely imaginary. Since the postquench energy $h^{(1)}_k$ is purely imaginary, the corresponding Fisher zeros are located on the real axis of the complex time plane, as shown in Fig.~\ref{Fig4}(b). For each fixed $n$, $z_n(k)$ is continuous and finite over the Brillouin zone, so the corresponding Fisher zeros form a finite line segment on the real axis. The branches shown in Fig.~\ref{Fig4}(b) form separated segments. None of these Fisher zero lines reaches the imaginary axis, and the rate function in Fig.~\ref{Fig4}(c) remains smooth and shows no nonanalytic cusp, indicating that no DQPT occurs.

\begin{figure}
	\centering
	{\includegraphics[width=1\columnwidth]{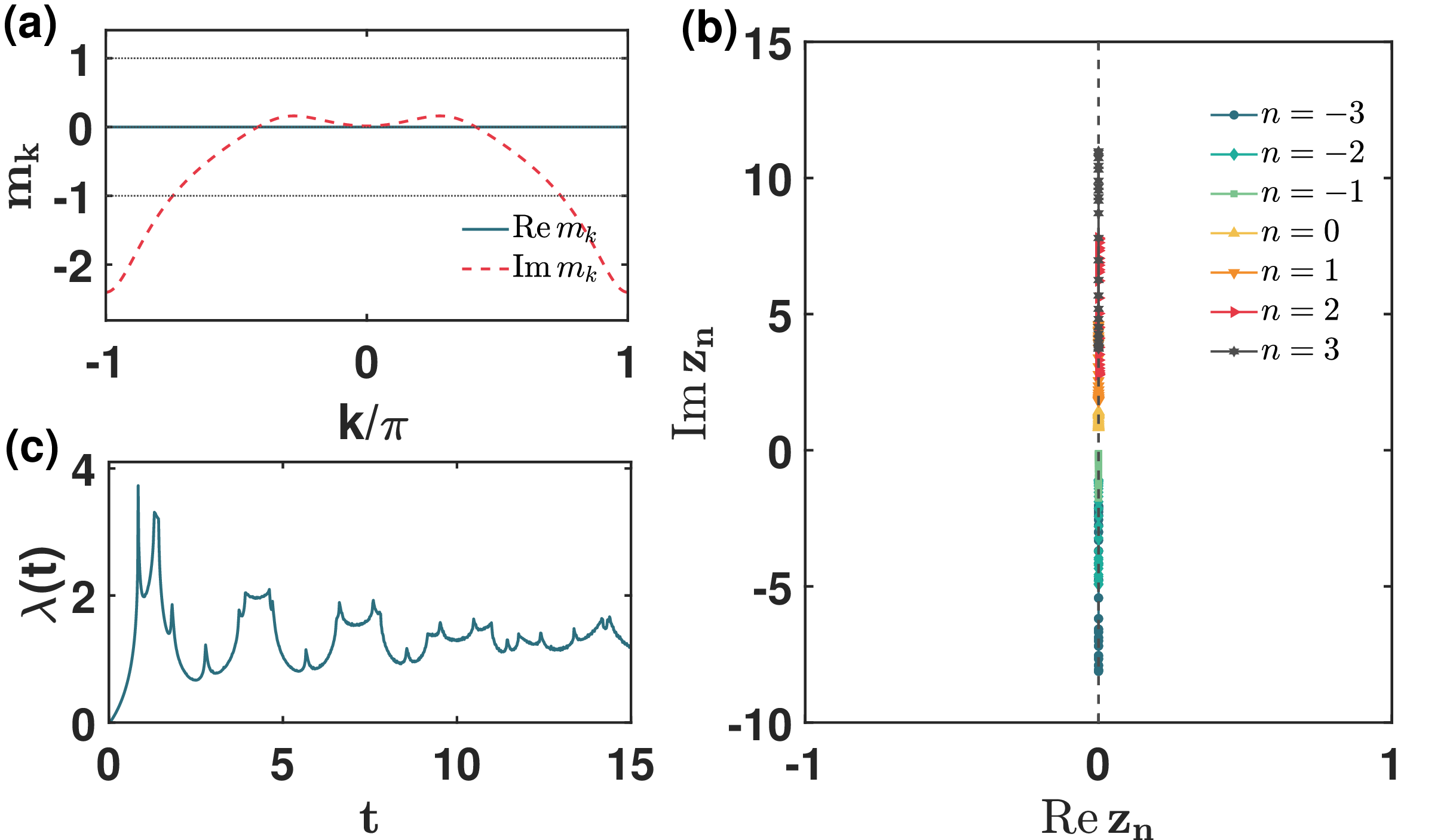}}
	\caption{Results for the BU quench from $(\phi_i,\gamma_i)=(\pi,3.5)$ to $(\phi_f,\gamma_f)=(-\pi,0.5)$. (a) The real and imaginary parts of $m_k$ as functions of $k$, where the blue solid and red dashed lines denote $\operatorname{Re}m_k$ and $\operatorname{Im}m_k$, respectively. (b) Fisher zeros $z_n(k)$ in the complex time plane for different $n$, with different colors and symbols representing different branches. The vertical dashed line marks $\operatorname{Re}z=0$. (c) The rate function $\lambda(t)$ as a function of time $t$.
	} 
	\label{Fig5}
\end{figure}
For the BU quench from $(\phi_i,\gamma_i)=(\pi,3.5)$ to $(\phi_f,\gamma_f)=(-\pi,0.5)$, the initial spectrum is purely imaginary, while the final spectrum is purely real, so $m_k$ is also purely imaginary, as shown in Fig.~\ref{Fig5}(a). Consequently, $Q_k=\tanh^{-1}m_k$ is purely imaginary. Since the postquench energy $h^{(1)}_k$ is real, the Fisher zeros can be reduced to
\begin{equation}
z_n\left( k \right) =i\frac{a-Q_{k,\mathrm{I}}\left( k \right)}{h_{k,\mathrm{R}}^{(1)}\left( k \right)}
\end{equation}
showing that $\operatorname{Re}z_n\left( k \right) =0$ is automatically satisfied for every $k$. Unlike the usual one-dimensional case, where only several isolated momentum modes satisfy the critical-momentum condition, all momentum modes $k$ in the present case satisfy the critical-momentum condition. Therefore, as shown in Fig.~\ref{Fig5}(b), all the Fisher zeros lie on the imaginary axis. For a fixed $n$, these Fisher zeros cover a continuous time interval $[t_{c}^{-},t_{c}^{+}]$. 

\begin{figure}
	\centering
	{\includegraphics[width=1\columnwidth]{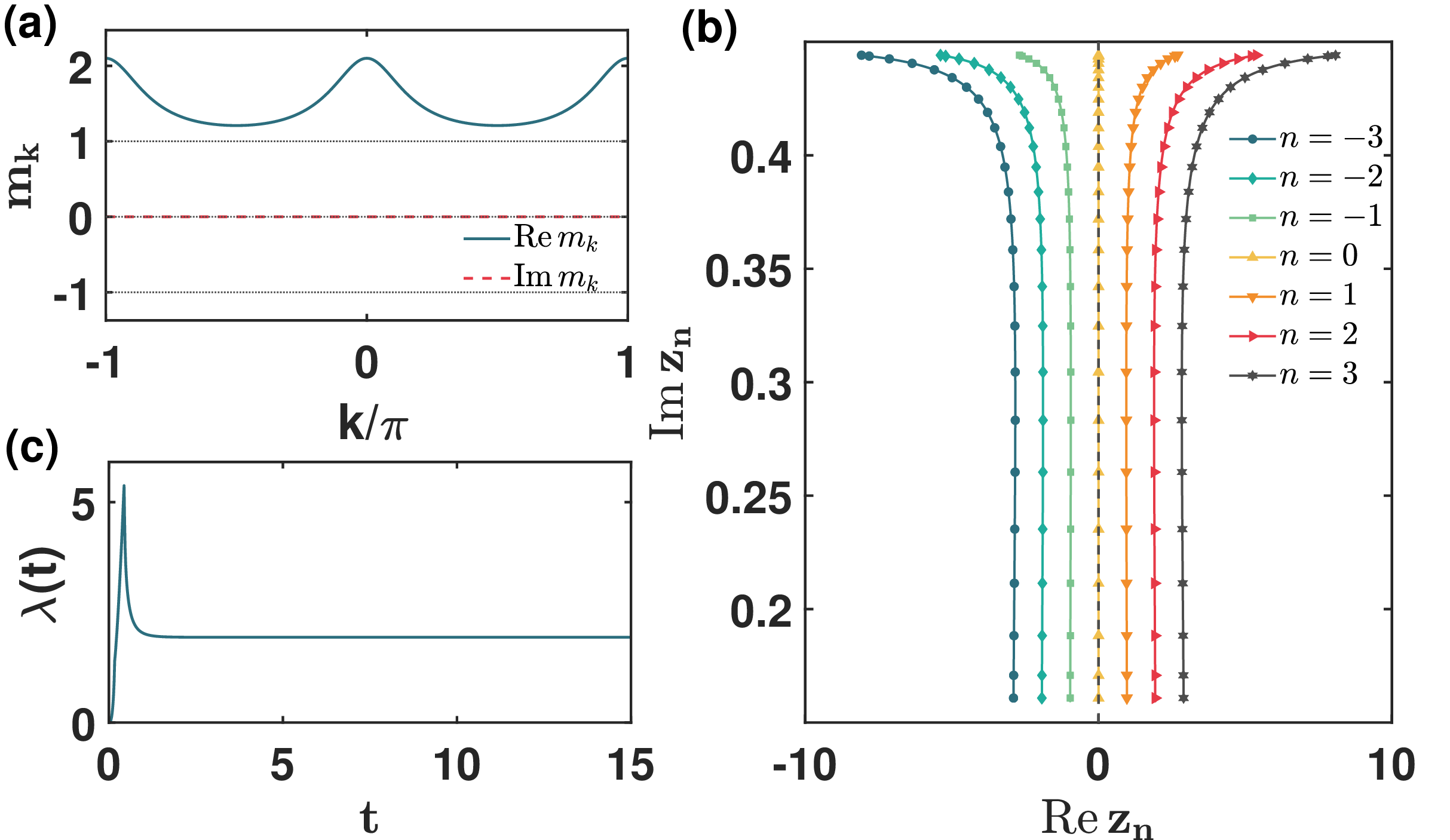}}
	\caption{Results for the BB quench from $(\phi_i,\gamma_i)=(\pi,3.5)$ to $(\phi_f,\gamma_f)=(0,3.5)$. (a) The real and imaginary parts of $m_k$ as functions of $k$, where the blue solid and red dashed lines denote $\operatorname{Re}m_k$ and $\operatorname{Im}m_k$, respectively. (b) Fisher zeros $z_n(k)$ in the complex time plane for different $n$, with different colors and symbols representing different branches. The vertical dashed line marks $\operatorname{Re}z=0$. (c) The rate function $\lambda(t)$ as a function of time $t$.
	} 
	\label{Fig6}
\end{figure}
For the BB quench, the Peierls phase is quenched from $\phi_i=\pi$ to $\phi_f=0$ while keeping $\gamma_i=\gamma_f=3.5$ fixed. For these parameters, both the initial and final spectra are purely imaginary, while $m_k$ is real and remains $m_k>1$, as shown in Fig.~\ref{Fig6}(a). Since the postquench energy is purely imaginary, $h_k^{(1)}=i h_{k,\mathrm{I}}^{(1)}$, the Fisher zeros satisfy
\begin{equation}
z_n=\frac{a_n-Q_{k,\mathrm{I}}}{h_{k,\mathrm{I}}^{(1)}}+i\frac{Q_{k,\mathrm{R}}}{h_{k,\mathrm{I}}^{(1)}}=\frac{\pi \left( n-l \right) }{h_{k,\mathrm{I}}^{(1)}}+i\frac{Q_{k,\mathrm{R}}}{h_{k,\mathrm{I}}^{(1)}}. 
\end{equation}
Therefore, the Fisher zero branch with $n=0$ lies on the imaginary axis for the principal branch $l=0$, as shown in Fig.~\ref{Fig6}(b), whereas the branches with $n>0$ and $n<0$ are distributed on the right and left sides of the imaginary axis, respectively.

\begin{figure}
	\centering
	{\includegraphics[width=1\columnwidth]{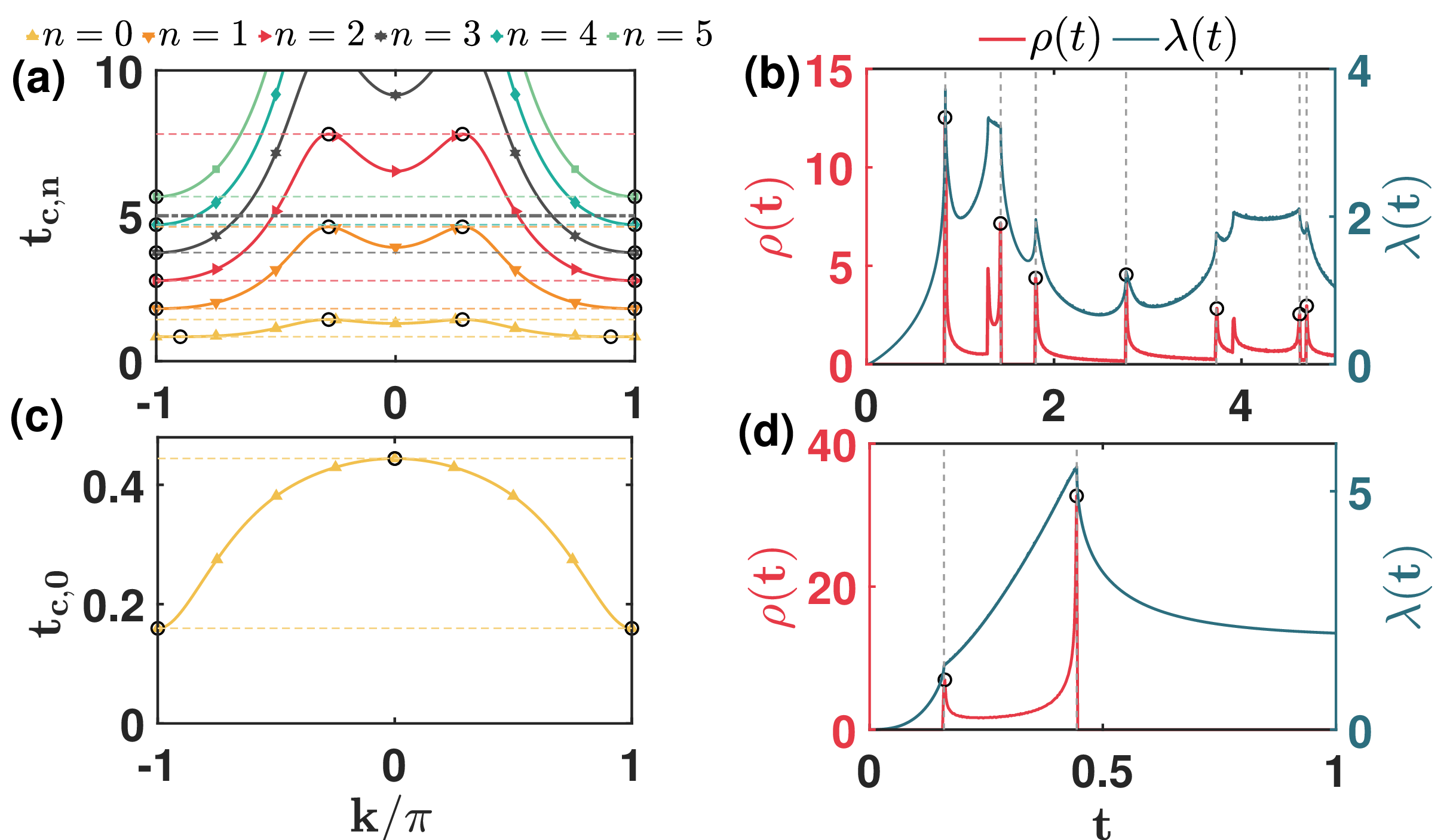}}
	\caption{Panels (a) and (c) show the critical times $t_{c,n}(k)=\operatorname{Im}z_n(k)$ of the Fisher zeros on the imaginary axis as functions of momentum $k$ for the BU and BB quenches, respectively. The colored horizontal dashed lines and circles mark the boundaries of the critical region for each branch. The black horizontal dashed line at $t=5$ in panel (a) indicates that only results for $0\leq t\leq5$ are shown in panel (b). Panels (b) and (d) show the density of Fisher zeros $\rho(t)$ (red curves) and the rate function $\lambda(t)$ (blue curves) for the corresponding quenches. The gray vertical dashed lines mark the nonanalytic points of the rate function, which coincide with the singularities of the Fisher zero density.
	} 
	\label{Fig7}
\end{figure}

Figs.~\ref{Fig7}(a) and \ref{Fig7}(c) show $t_{c,n}(k)=\operatorname{Im}z_n(k)$ as a function of momentum $k$ for the Fisher zero branches on the imaginary axis in the BU and BB quenches. Each branch covers a continuous interval $[t_c^{-},t_c^{+}]$ on the imaginary axis, and the boundary times correspond to singularities in the Fisher zero line density $\rho(t)$ shown in Figs.~\ref{Fig7}(b) and \ref{Fig7}(d). For the BU quench, additional density singularities occur not only at their boundaries but also inside these intervals. As shown in Figs.~\ref{Fig7}(b) and \ref{Fig7}(d), all these density singularities correspond to nonanalytic points of the rate function $\lambda(t)$, showing that DQPTs can occur regardless of whether the system crosses a PT phase boundary. Therefore, the line density of Fisher zeros not only describes their distribution within the critical regions but also allows us to identify the DQPT critical times within the regions continuously covered by Fisher zeros.

\subsection{Quench dynamics with exceptional points in the postquench Hamiltonian}\label{3.3}

All the cases of UU, UB, BU, and BB discussed above involve prequench and postquench Hamiltonians which can be diagonalized for all momenta. However, for the cases of UP and BP, where the postquench Hamiltonian is in the partially broken PT phase, the momentum regions with real and purely imaginary postquench spectra are separated by isolated momenta at which $H_k^{(1)}$ reaches EPs. At such a momentum, the two eigenvalues and their corresponding eigenvectors coalesce, and $H_k^{(1)}$ cannot be diagonalized. Therefore, before analyzing the UP and BP quenches, we first determine the postquench dynamics at these EPs using the Jordan decomposition, which reads\cite{Heiss_2012}
\begin{eqnarray}
H_{\mathrm{EP}}=S J_{\mathrm{EP}} S^{-1}=
S\left( \begin{matrix}
   E_{\text{EP}}&		1\\
	0&		E_{\text{EP}}\\
	\end{matrix} \right)       S^{-1}. 
\end{eqnarray}
For the traceless two-band Hamiltonian, the coalescence condition requires $E_{\mathrm{EP}}=0$, the Jordan block therefore reduces to
\begin{eqnarray}
	J_{\text{EP}}=\left( \begin{matrix}
		0&		1\\
		0&		0\\
	\end{matrix} \right) .
\end{eqnarray}
$H_{\mathrm{EP}}$ is related to $J_{\mathrm{EP}}$ by a similarity transformation, so that $ H_\mathrm{EP}^{2}=0$ and $H_{\mathrm EP}\neq0 $. The time evolution operator at the EP can thus be evaluated as 
\begin{eqnarray}
e^{-iH_\mathrm{EP}t}=I-iH_\mathrm{EP}t.
\end{eqnarray}
If the prequench Hamiltonian is away from EPs and the postquench Hamiltonian reaches an EP, the coefficients entering the Loschmidt echo in Eq.~(\ref{LE}) can be expressed as
\begin{eqnarray}
	&a_{\mathrm{EP}}(t) =-it\left< \tilde{u}_{k+}^{(0)}|H_{\mathrm{EP}}^{(1)}|u_{k-}^{(0)} \right>    \nonumber  \\
	&G_k(t) =b_\mathrm{EP}(t) =1-it\left< \tilde{u}_{k-}^{(0)}|H_{\mathrm{EP}}^{(1)}|u_{k-}^{(0)} \right> .
\end{eqnarray}
Although the postquench Hamiltonian is not diagonalizable at the EPs, its time-evolution operator remains well defined and can be obtained from the Jordan form. Therefore, the biorthogonal Loschmidt amplitude and the Loschmidt echo can be evaluated at EPs. 

\begin{figure}
	\centering
	{\includegraphics[width=1\columnwidth]{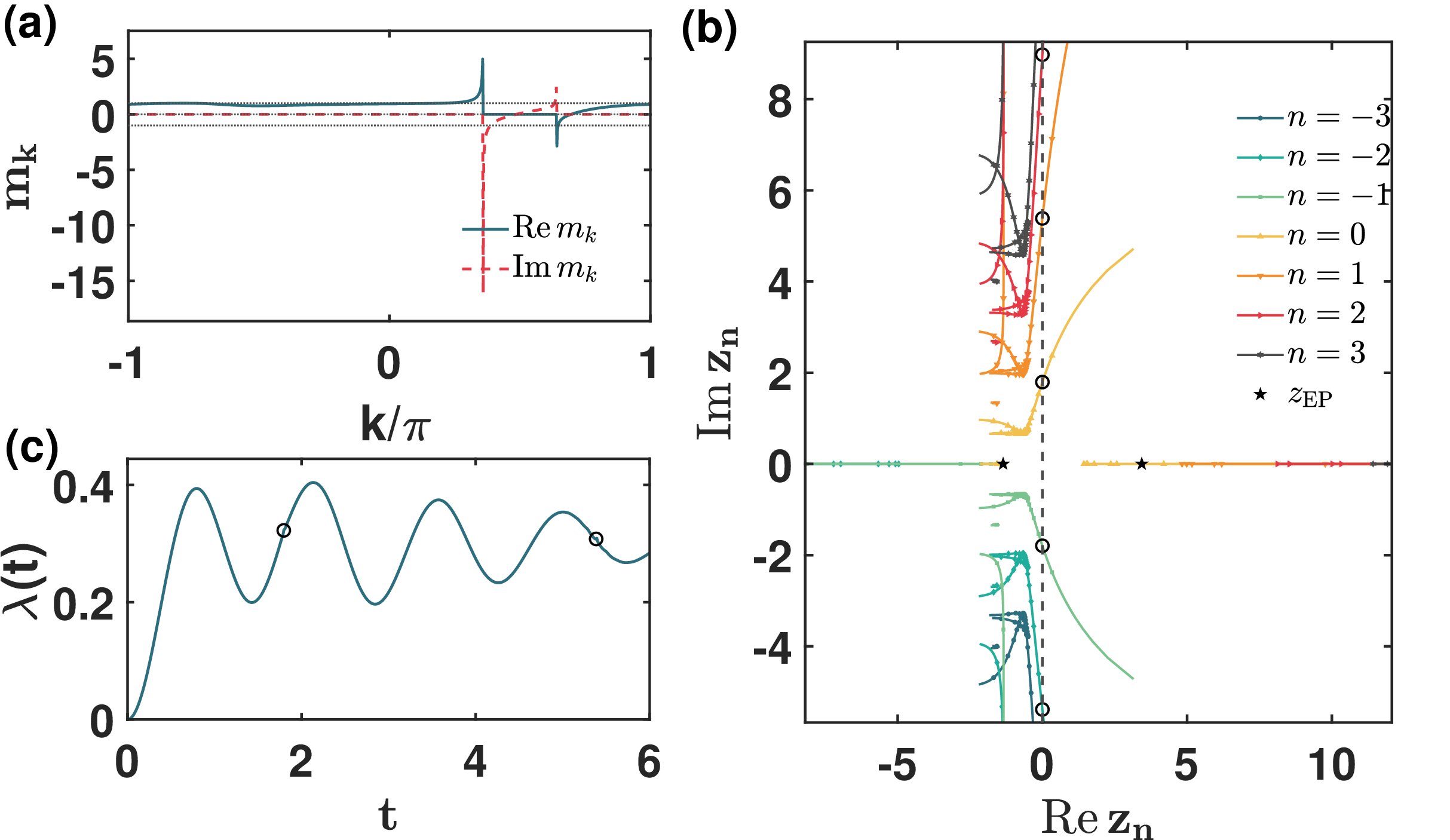}}
	\caption{Results for the UP quench from $(\phi_i,\gamma_i)=(0,0.5)$ to $(\phi_f,\gamma_f)=(0.5\pi,1)$. (a) The real and imaginary parts of $m_k$ as functions of $k$, where the blue solid and red dashed lines denote $\operatorname{Re}m_k$ and $\operatorname{Im}m_k$, respectively. (b) Fisher zeros $z_n(k)$ in the complex time plane for different $n$, with different colors and symbols representing different branches. In the thermodynamic limit, the vertical dashed line marks $\operatorname{Re}z=0$, and the black stars denote the Fisher zeros $z_{\rm EP}$ evaluated at the exceptional momenta $k_{\rm EP}$. (c) The rate function $\lambda(t)$ as a function of time $t$. The circles in panels (b) and (c) mark the critical times at which DQPTs occur.
	} 
	\label{Fig8}
\end{figure}

For the UP quench from $(\phi_i,\gamma_i)=(0,0.5)$ to $(\phi_f,\gamma_f)=(0.5\pi,1)$, the initial spectrum is purely real, while the postquench Hamiltonian is in the partially broken PT phase, and the Fisher zero lines can exhibit features of both the UU and UB cases. As shown in Fig.~\ref{Fig8}(a), $m_k$ is either purely real or purely imaginary, depending on the spectral region of the postquench Hamiltonian. The sharp changes of $m_k$ in Fig.~\ref{Fig8}(a) occur as $k$ crosses the exceptional momenta $k=k_{\mathrm{EP}}$, which separate the momentum regions with real and purely imaginary postquench spectra. At the EPs, $h_{k_\mathrm{EP}}^{(1)}=0$, so the usual Fisher zero expression is not applicable and the corresponding zeros $z_{\rm EP}$ are evaluated separately using the Jordan form. As shown in Fig.~\ref{Fig8}(b), when $m_k$ is real, it can cross $m_k=\pm1$, leading to the extension and separation of the Fisher zero lines, and when $m_k$ is purely imaginary, the Fisher zeros lie on the real axis. Some Fisher zero lines cross the imaginary axis, and the corresponding critical times agree with the nonanalytic points of the rate function $\lambda(t)$ in Fig.~\ref{Fig8}(c), indicating the occurrence of DQPTs. 

\begin{figure}
	\centering
	{\includegraphics[width=1\columnwidth]{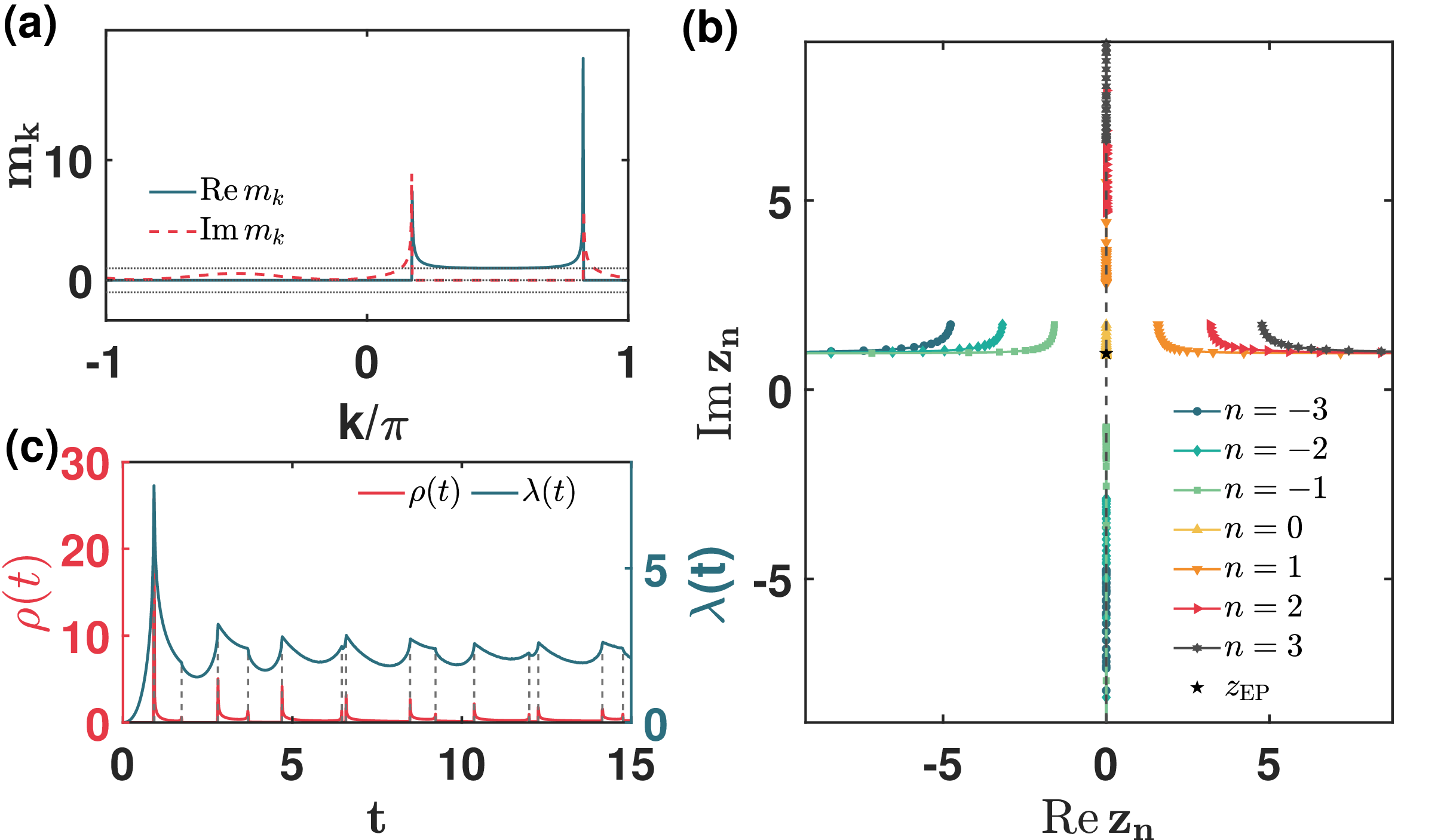}}
	\caption{Results for the BP quench from $(\phi_i,\gamma_i)=(0.5\pi,3.5)$ to $(\phi_f,\gamma_f)=(0.5\pi,2)$. (a) The real and imaginary parts of $m_k$ as functions of $k$, where the blue solid and red dashed lines denote $\operatorname{Re}m_k$ and $\operatorname{Im}m_k$, respectively. (b) Fisher zeros $z_n(k)$ in the complex time plane for different $n$, with different colors and symbols representing different branches. In the thermodynamic limit, the vertical dashed line marks $\operatorname{Re}z=0$, and the black stars denote the Fisher zeros $z_{\rm EP}$ evaluated at the exceptional momenta $k_{\rm EP}$. (c)The density of Fisher zeros $\rho(t)$ (red curves) and the rate function $\lambda(t)$ (blue curves). 
	} 
	\label{Fig9}
\end{figure}

For the BP quench from $(\phi_i,\gamma_i)=(0.5\pi,3.5)$ to $(\phi_f,\gamma_f)=(0.5\pi,2)$, the prequench spectrum is purely imaginary, in contrast to the UP case. As shown in Fig.~\ref{Fig9}(a), $m_k$ is purely imaginary when the postquench spectrum is real and becomes real when the postquench spectrum is purely imaginary. The Fisher zero lines exhibit the structures discussed for both the BU and BB cases. In the former region, the Fisher zeros lie on the imaginary axis and form finite intervals, while in the latter region, the $n=0$ Fisher zero branch lies on the imaginary axis, and the other branches lie to the left or right of the imaginary axis, as shown in Fig.~\ref{Fig9}(b). The Fisher zero distribution on the imaginary axis gives the corresponding critical times, which are consistent with the nonanalytic points of the rate function $\lambda(t)$ shown in Fig.~\ref{Fig9}(c).
\section{conclusion}\label{sec4} 
In this work, we explain the origin of the anomalous behavior of non-Hermitian Fisher zero lines by analyzing the dot product $m_k$ of the normalized complex Bloch vectors before and after quenching. Away from exceptional points, as real $m_k\rightarrow\pm1$, the real part of $Q_k$ diverges, causing the Fisher zero lines to extend to infinity and appear separated within the finite range shown in the figure. For real $m_k$ with $|m_k|>1$, the Fisher zero branch with $n=l$ is located on the real or imaginary axis of the complex time plane for a purely real or purely imaginary postquench spectrum, respectively. For purely imaginary $m_k$, these two types of spectra instead cause all Fisher zero branches to lie on the imaginary or the real axis, respectively. When Fisher zeros continuously cover the real-time axis, we introduce the line density of Fisher zeros and find that the locations where the density of Fisher zeros becomes singular correspond to the nonanalytic points of the rate function, thereby identifying the DQPT critical times from the continuous critical regions. This connects the anomalous Fisher zero structures with dynamical critical behavior in the thermodynamic limit. We verify these results in a generalized non-Hermitian SSH model with a Peierls phase. In particular, a Peierls phase quench can induce biorthogonal DQPTs within the fully broken PT phase, showing that crossing a PT phase boundary is not necessary for DQPTs to occur. At exceptional points of the postquench Hamiltonian, we evaluate the dynamics using the Jordan form because the conventional Fisher zero expression cannot be applied directly. Our research shows that $m_k=\hat{\mathbf{h}}_0\cdot\hat{\mathbf{h}}_1$ remains an important physical quantity in non-Hermitian systems even though the Hermitian critical momentum condition is no longer universal. Its broader range of values plays an important role in understanding anomalous Fisher zero structures, and our results offer a clearer physical picture of dynamical critical behavior in non-Hermitian systems.

\begin{acknowledgments}
This work was supported by National Natural Science Foundation of China (NSFC) under grant No. 11475037, No. 11574041.\\
\end{acknowledgments}

\section*{Data Availability}
All the data that support the plots and other findings of this study are available from the corresponding author upon reasonable request.

\bibliography{references}

@article{PhysRevLett.110.135704,
	title = {Dynamical Quantum Phase Transitions in the Transverse-Field Ising Model},
	author = {Heyl, M. and Polkovnikov, A. and Kehrein, S.},
	journal = {Phys. Rev. Lett.},
	volume = {110},
	issue = {13},
	pages = {135704},
	numpages = {5},
	year = {2013},
	month = {Mar},
	publisher = {American Physical Society},
	doi = {10.1103/PhysRevLett.110.135704},
	url = {https://link.aps.org/doi/10.1103/PhysRevLett.110.135704}
}

@article{heyl2018dynamical,
	title={Dynamical quantum phase transitions: a review},
	author={Heyl, Markus},
	journal={Reports on Progress in Physics},
	volume={81},
	number={5},
	pages={054001},
	year={2018},
	month = {apr},
	publisher = {IOP Publishing},
	doi = {10.1088/1361-6633/aaaf9a},
	url = {https://doi.org/10.1088/1361-6633/aaaf9a},
}

@article{budich2016dynamical,
	title = {Dynamical topological order parameters far from equilibrium},
	author = {Budich, Jan Carl and Heyl, Markus},
	journal = {Phys. Rev. B},
	volume = {93},
	issue = {8},
	pages = {085416},
	numpages = {7},
	year = {2016},
	month = {Feb},
	publisher = {American Physical Society},
	doi = {10.1103/PhysRevB.93.085416},
	url = {https://link.aps.org/doi/10.1103/PhysRevB.93.085416}
}

@article{Zvyagin,
	author = {Zvyagin, A. A.},
	title = {Dynamical quantum phase transitions (Review Article)},
	journal = {Low Temperature Physics},
	volume = {42},
	number = {11},
	pages = {971-994},
	year = {2016},
	month = {11},
	issn = {1063-777X},
	doi = {10.1063/1.4969869},
	url = {https://doi.org/10.1063/1.4969869},
}

@article{Heyl_2019,
	doi = {10.1209/0295-5075/125/26001},
	url = {https://doi.org/10.1209/0295-5075/125/26001},
	year = {2019},
	month = {feb},
	publisher = {EDP Sciences, IOP Publishing and Società Italiana di Fisica},
	volume = {125},
	number = {2},
	pages = {26001},
	author = {Heyl, Markus},
	title = {Dynamical quantum phase transitions: A brief survey},
	journal = {Europhysics Letters}
}

@article{bhattacharya2017emergent,
	title = {Emergent topology and dynamical quantum phase transitions in two-dimensional closed quantum systems},
	author = {Bhattacharya, Utso and Dutta, Amit},
	journal = {Phys. Rev. B},
	volume = {96},
	issue = {1},
	pages = {014302},
	numpages = {6},
	year = {2017},
	month = {Jul},
	publisher = {American Physical Society},
	doi = {10.1103/PhysRevB.96.014302},
	url = {https://link.aps.org/doi/10.1103/PhysRevB.96.014302}
}

@article{vajna2014disentangling,
	title = {Disentangling dynamical phase transitions from equilibrium phase transitions},
	author = {Vajna, Szabolcs and D{\'o}ra, Bal{\'a}zs},
	journal = {Phys. Rev. B},
	volume = {89},
	issue = {16},
	pages = {161105(R)},
	numpages = {5},
	year = {2014},
	month = {Apr},
	publisher = {American Physical Society},
	doi = {10.1103/PhysRevB.89.161105},
	url = {https://link.aps.org/doi/10.1103/PhysRevB.89.161105}
}

@article{vajna2015topological,
	title = {Topological classification of dynamical phase transitions},
	author = {Vajna, Szabolcs and D{\'o}ra, Bal{\'a}zs},
	journal = {Phys. Rev. B},
	volume = {91},
	issue = {15},
	pages = {155127},
	numpages = {5},
	year = {2015},
	month = {Apr},
	publisher = {American Physical Society},
	doi = {10.1103/PhysRevB.91.155127},
	url = {https://link.aps.org/doi/10.1103/PhysRevB.91.155127}
}

@article{wong2024entanglement,
title = {Entanglement in quenched extended Su-Schrieffer-Heeger model with anomalous dynamical quantum phase transitions},
author = {Wong, Cheuk Yiu and Hui, Tsz Hin and Sacramento, P. D. and Yu, Wing Chi},
journal = {Phys. Rev. B},
volume = {110},
issue = {5},
pages = {054312},
numpages = {17},
year = {2024},
month = {Aug},
publisher = {American Physical Society},
doi = {10.1103/PhysRevB.110.054312},
url = {https://link.aps.org/doi/10.1103/PhysRevB.110.054312}
}

@article{Chenshu,
	title = {Relation between equilibrium quantum phase transitions and dynamical quantum phase transitions in two-band systems},
	author = {Zeng, Yumeng and Chen, Shu},
	journal = {Phys. Rev. B},
	volume = {112},
	issue = {6},
	pages = {064307},
	numpages = {9},
	year = {2025},
	month = {Aug},
	publisher = {American Physical Society},
	doi = {10.1103/srx7-cpl4},
	url = {https://link.aps.org/doi/10.1103/srx7-cpl4}
}

@article{sharma2016slow,
	title = {Slow quenches in a quantum Ising chain: Dynamical phase transitions and topology},
	author = {Sharma, Shraddha and Divakaran, Uma and Polkovnikov, Anatoli and Dutta, Amit},
	journal = {Phys. Rev. B},
	volume = {93},
	issue = {14},
	pages = {144306},
	numpages = {9},
	year = {2016},
	month = {Apr},
	publisher = {American Physical Society},
	doi = {10.1103/PhysRevB.93.144306},
	url = {https://link.aps.org/doi/10.1103/PhysRevB.93.144306}
}

@article{PhysRevB.95.184307,
	title = {Interconnections between equilibrium topology and dynamical quantum phase transitions in a linearly ramped Haldane model},
	author = {Bhattacharya, Utso and Dutta, Amit},
	journal = {Phys. Rev. B},
	volume = {95},
	issue = {18},
	pages = {184307},
	numpages = {13},
	year = {2017},
	month = {May},
	publisher = {American Physical Society},
	doi = {10.1103/PhysRevB.95.184307},
	url = {https://link.aps.org/doi/10.1103/PhysRevB.95.184307}
}

@article{zamani2024scaling,
	title = {Scaling and universality at ramped quench dynamical quantum phase transitions},
	author = {Zamani, Sara and Naji, J and Jafari, R and Langari, A},
	journal = {Journal of Physics: Condensed Matter},
	year = {2024},
	month = {jun},
	volume = {36},
	number = {35},
	pages = {355401},
	journal = {Journal of Physics: Condensed Matter},
	doi = {10.1088/1361-648X/ad4df9},
	url = {https://doi.org/10.1088/1361-648X/ad4df9}
}

@article{baghran2024competition,
	title = {Competition of long-range interactions and noise at a ramped quench dynamical quantum phase transition: The case of the long-range pairing Kitaev chain},
	author = {Baghran, R. and Jafari, R. and Langari, A.},
	journal = {Phys. Rev. B},
	volume = {110},
	issue = {6},
	pages = {064302},
	numpages = {11},
	year = {2024},
	month = {Aug},
	publisher = {American Physical Society},
	doi = {10.1103/PhysRevB.110.064302},
	url = {https://link.aps.org/doi/10.1103/PhysRevB.110.064302}
}

@article{PhysRevB.109.L180303,
	title = {Dynamical quantum phase transitions following a noisy quench},
	author = {Jafari, R. and Langari, A. and Eggert, S. and Johannesson, Henrik},
	journal = {Phys. Rev. B},
	volume = {109},
	issue = {18},
	pages = {L180303},
	numpages = {7},
	year = {2024},
	month = {May},
	publisher = {American Physical Society},
	doi = {10.1103/PhysRevB.109.L180303},
	url = {https://link.aps.org/doi/10.1103/PhysRevB.109.L180303}
}

@article{cao2024exploring,
	title = {Exploring dynamical phase transitions in the $XY$ chain through a linear quench: Early and long-term perspectives},
	author = {Cao, Kaiyuan and Hou, Hongsheng and Tong, Peiqing},
	journal = {Phys. Rev. A},
	volume = {110},
	issue = {4},
	pages = {042209},
	numpages = {12},
	year = {2024},
	month = {Oct},
	publisher = {American Physical Society},
	doi = {10.1103/PhysRevA.110.042209},
	url = {https://link.aps.org/doi/10.1103/PhysRevA.110.042209}
}

@article{32c2-z9hf,
	title = {Fate of dynamical quantum phase transitions from sudden to slow quench limit},
	author = {Zhang, Xiang and Li, Fuxiang},
	journal = {Phys. Rev. B},
	volume = {114},
	issue = {1},
	pages = {014320},
	numpages = {11},
	year = {2026},
	month = {Jul},
	publisher = {American Physical Society},
	doi = {10.1103/32c2-z9hf},
	url = {https://link.aps.org/doi/10.1103/32c2-z9hf}
}

@article{PhysRevB.106.045410,
	title = {Nonlinear current and dynamical quantum phase transitions in the flux-quenched Su-Schrieffer-Heeger model},
	author = {Rossi, Lorenzo and Dolcini, Fabrizio},
	journal = {Phys. Rev. B},
	volume = {106},
	issue = {4},
	pages = {045410},
	numpages = {8},
	year = {2022},
	month = {Jul},
	publisher = {American Physical Society},
	doi = {10.1103/PhysRevB.106.045410},
	url = {https://link.aps.org/doi/10.1103/PhysRevB.106.045410}
}

@article{nie2024flux,
	title={Flux-quench induced dynamical quantum phase transitions in an extended XY spin-chain},
	author={Nie, Wen-Hui and Zhang, Mei-Yu and Wang, Lin-Cheng},
	journal={Physica A: Statistical Mechanics and its Applications},
	volume={653},
	pages={130110},
	year={2024},
	doi = {https://doi.org/10.1016/j.physa.2024.130110},
	url = {https://www.sciencedirect.com/science/article/pii/S0378437124006198}
}

@article{zhang2025phase,
	title = {Phase quench effects on dynamical quantum phase transitions in the Su-Schrieffer-Heeger model},
	author = {Mei-Yu Zhang and Wen-Hui Nie and Lin-Cheng Wang},
	journal = {Physics Letters A},
	volume = {566},
	pages = {131193},
	year = {2026},
	issn = {0375-9601},
	doi = {https://doi.org/10.1016/j.physleta.2025.131193},
	url = {https://www.sciencedirect.com/science/article/pii/S0375960125009739}
}

@article{nie2026Floquet,
	title = {Floquet dynamical quantum phase transitions in periodically flux-quenched systems},
	journal = {Physica A: Statistical Mechanics and its Applications},
	volume = {693},
	pages = {131568},
	year = {2026},
	issn = {0378-4371},
	doi = {https://doi.org/10.1016/j.physa.2026.131568},
	url = {https://www.sciencedirect.com/science/article/pii/S0378437126003043},
	author = {Wen-Hui Nie and Mei-Yu Zhang and Lin-Cheng Wang and Chong Li}
}

@article{PhysRevB.100.085308,
	title = {Floquet dynamical quantum phase transitions},
	author = {Yang, Kai and Zhou, Longwen and Ma, Wenchao and Kong, Xi and Wang, Pengfei and Qin, Xi and Rong, Xing and Wang, Ya and Shi, Fazhan and Gong, Jiangbin and Du, Jiangfeng},
	journal = {Phys. Rev. B},
	volume = {100},
	issue = {8},
	pages = {085308},
	numpages = {11},
	year = {2019},
	month = {Aug},
	publisher = {American Physical Society},
	doi = {10.1103/PhysRevB.100.085308},
	url = {https://link.aps.org/doi/10.1103/PhysRevB.100.085308}
}

@article{PhysRevB.102.144306,
	title = {Floquet dynamical quantum phase transition in the extended XY model: Nonadiabatic to adiabatic topological transition},
	author = {Zamani, Sara and Jafari, R. and Langari, A.},
	journal = {Phys. Rev. B},
	volume = {102},
	issue = {14},
	pages = {144306},
	numpages = {12},
	year = {2020},
	month = {Oct},
	publisher = {American Physical Society},
	doi = {10.1103/PhysRevB.102.144306},
	url = {https://link.aps.org/doi/10.1103/PhysRevB.102.144306}
}

@article{zhou2021floquet,
	title={Floquet dynamical quantum phase transitions in periodically quenched systems},
	author={Zhou, Longwen and Du, Qianqian},
	journal={Journal of Physics: Condensed Matter},
	volume={33},
	number={34},
	pages={345403},
	year={2021},
	month = {jul},
	publisher = {IOP Publishing},
	doi = {10.1088/1361-648X/ac0b60},
	url = {https://doi.org/10.1088/1361-648X/ac0b60}
}

@article{PhysRevA.103.012204,
	title = {Floquet dynamical phase transition and entanglement spectrum},
	author = {Jafari, R. and Akbari, Alireza},
	journal = {Phys. Rev. A},
	volume = {103},
	issue = {1},
	pages = {012204},
	numpages = {8},
	year = {2021},
	month = {Jan},
	publisher = {American Physical Society},
	doi = {10.1103/PhysRevA.103.012204},
	url = {https://link.aps.org/doi/10.1103/PhysRevA.103.012204}
}

@article{luan2022floquet,
	title={Floquet dynamical quantum phase transitions of the XY spin-chain under periodic quenching},
	author={Luan, Li-Na and Zhang, Mei-Yu and Wang, LC},
	journal={Physica A: Statistical Mechanics and its Applications},
	volume={604},
	pages={127866},
	year={2022},
	publisher={Elsevier},
	doi = {https://doi.org/10.1016/j.physa.2022.127866},
	url = {https://www.sciencedirect.com/science/article/pii/S0378437122005593}
}

@article{PhysRevB.106.094314,
	title = {Engineering Floquet dynamical quantum phase transitions},
	author = {Naji, J. and Jafari, R. and Zhou, Longwen and Langari, A.},
	journal = {Phys. Rev. B},
	volume = {106},
	issue = {9},
	pages = {094314},
	numpages = {10},
	year = {2022},
	month = {Sep},
	publisher = {American Physical Society},
	doi = {10.1103/PhysRevB.106.094314},
	url = {https://link.aps.org/doi/10.1103/PhysRevB.106.094314}
}

@article{PhysRevB.105.094311,
	title = {Floquet dynamical quantum phase transitions under synchronized periodic driving},
	author = {Jafari, R. and Akbari, Alireza and Mishra, Utkarsh and Johannesson, Henrik},
	journal = {Phys. Rev. B},
	volume = {105},
	issue = {9},
	pages = {094311},
	numpages = {13},
	year = {2022},
	month = {Mar},
	publisher = {American Physical Society},
	doi = {10.1103/PhysRevB.105.094311},
	url = {https://link.aps.org/doi/10.1103/PhysRevB.105.094311}
}

@article{PhysRevB.96.180303,
	title = {Mixed state dynamical quantum phase transitions},
	author = {Bhattacharya, Utso and Bandyopadhyay, Souvik and Dutta, Amit},
	journal = {Phys. Rev. B},
	volume = {96},
	issue = {18},
	pages = {180303(R)},
	numpages = {5},
	year = {2017},
	month = {Nov},
	publisher = {American Physical Society},
	doi = {10.1103/PhysRevB.96.180303},
	url = {https://link.aps.org/doi/10.1103/PhysRevB.96.180303}
}

@article{PhysRevB.96.180304,
	title = {Dynamical topological quantum phase transitions for mixed states},
	author = {Heyl, M. and Budich, J. C.},
	journal = {Phys. Rev. B},
	volume = {96},
	issue = {18},
	pages = {180304(R)},
	numpages = {5},
	year = {2017},
	month = {Nov},
	publisher = {American Physical Society},
	doi = {10.1103/PhysRevB.96.180304},
	url = {https://link.aps.org/doi/10.1103/PhysRevB.96.180304}
}

@article{cao2025,
	title = {Exploring dynamical quantum phase transitions from pure states to mixed states through extended Su-Schrieffer-Heeger models},
	author = {Cao, Kaiyuan and Zhang, Tianren and Jiang, Xiang-Ping and Wang, Jian},
	journal = {Phys. Rev. A},
	volume = {112},
	issue = {4},
	pages = {042217},
	numpages = {14},
	year = {2025},
	month = {Oct},
	publisher = {American Physical Society},
	doi = {10.1103/5289-8kz5},
	url = {https://link.aps.org/doi/10.1103/5289-8kz5}
}

@article{PhysRevB.98.134310,
	title = {Dynamical quantum phase transition for mixed states in open systems},
	author = {Lang, Haifeng and Chen, Yixin and Hong, Qiantan and Fan, Heng},
	journal = {Phys. Rev. B},
	volume = {98},
	issue = {13},
	pages = {134310},
	numpages = {7},
	year = {2018},
	month = {Oct},
	publisher = {American Physical Society},
	doi = {10.1103/PhysRevB.98.134310},
	url = {https://link.aps.org/doi/10.1103/PhysRevB.98.134310}
}

@article{PhysRevA.105.022220,
	title = {Dissipative Floquet dynamical quantum phase transition},
	author = {Naji, J. and Jafari, Masoud and Jafari, R. and Akbari, Alireza},
	journal = {Phys. Rev. A},
	volume = {105},
	issue = {2},
	pages = {022220},
	numpages = {11},
	year = {2022},
	month = {Feb},
	publisher = {American Physical Society},
	doi = {10.1103/PhysRevA.105.022220},
	url = {https://link.aps.org/doi/10.1103/PhysRevA.105.022220}
}

@article{PhysRevB.97.045147,
	title = {Fate of dynamical phase transitions at finite temperatures and in open systems},
	author = {Sedlmayr, N. and Fleischhauer, M. and Sirker, J.},
	journal = {Phys. Rev. B},
	volume = {97},
	issue = {4},
	pages = {045147},
	numpages = {8},
	year = {2018},
	month = {Jan},
	publisher = {American Physical Society},
	doi = {10.1103/PhysRevB.97.045147},
	url = {https://link.aps.org/doi/10.1103/PhysRevB.97.045147}
}

@article{gqv9-rfss,
	title = {Smearing of dynamical quantum phase transitions in dissipative free-fermion systems},
	author = {Parez, Gilles and Alba, Vincenzo},
	journal = {Phys. Rev. B},
	volume = {113},
	issue = {14},
	pages = {144310},
	numpages = {8},
	year = {2026},
	month = {Apr},
	publisher = {American Physical Society},
	doi = {10.1103/gqv9-rfss},
	url = {https://link.aps.org/doi/10.1103/gqv9-rfss}
}

@article{xu2026,
	title = {Generalized Loschmidt amplitudes and Uhlmann fidelity: Diagnostics of dynamical quantum phase transitions in a dissipative SSH chain},
	journal = {Physica A: Statistical Mechanics and its Applications},
	volume = {701},
	pages = {132022},
	year = {2026},
	issn = {0378-4371},
	doi = {https://doi.org/10.1016/j.physa.2026.132022},
	url = {https://www.sciencedirect.com/science/article/pii/S0378437126007582},
	author = {He-Guang Xu and Zheng Liu}
}

@article{PhysRevLett.119.080501,
	title = {Direct Observation of Dynamical Quantum Phase Transitions in an Interacting Many-Body System},
	author = {Jurcevic, P. and Shen, H. and Hauke, P. and Maier, C. and Brydges, T. and Hempel, C. and Lanyon, B. P. and Heyl, M. and Blatt, R. and Roos, C. F.},
	journal = {Phys. Rev. Lett.},
	volume = {119},
	issue = {8},
	pages = {080501},
	numpages = {5},
	year = {2017},
	month = {Aug},
	publisher = {American Physical Society},
	doi = {10.1103/PhysRevLett.119.080501},
	url = {https://link.aps.org/doi/10.1103/PhysRevLett.119.080501}
}

@article{Flaeschner2018,
	author  = {Fl{\"a}schner, N. and Vogel, D. and Tarnowski, M. and
	Rem, B. S. and L{\"u}hmann, D.-S. and Heyl, M. and
	Budich, J. C. and Mathey, L. and Sengstock, K. and
	Weitenberg, C.},
	title   = {Observation of dynamical vortices after quenches in a system with topology},
	journal = {Nature Physics},
	year    = {2018},
	volume  = {14},
	number  = {3},
	pages   = {265--268},
	doi     = {10.1038/s41567-017-0013-8},
	url     = {https://doi.org/10.1038/s41567-017-0013-8},
	issn    = {1745-2481}
}

@article{guo2019observation,
	title = {Observation of a Dynamical Quantum Phase Transition by a Superconducting Qubit Simulation},
	author = {Guo, Xue-Yi and Yang, Chao and Zeng, Yu and Peng, Yi and Li, He-Kang and Deng, Hui and Jin, Yi-Rong and Chen, Shu and Zheng, Dongning and Fan, Heng},
	journal = {Phys. Rev. Appl.},
	volume = {11},
	issue = {4},
	pages = {044080},
	numpages = {12},
	year = {2019},
	month = {Apr},
	publisher = {American Physical Society},
	doi = {10.1103/PhysRevApplied.11.044080},
	url = {https://link.aps.org/doi/10.1103/PhysRevApplied.11.044080}
}

@article{PhysRevLett.122.020501,
	title = {Simulating Dynamic Quantum Phase Transitions in Photonic Quantum Walks},
	author = {Wang, Kunkun and Qiu, Xingze and Xiao, Lei and Zhan, Xiang and Bian, Zhihao and Yi, Wei and Xue, Peng},
	journal = {Phys. Rev. Lett.},
	volume = {122},
	issue = {2},
	pages = {020501},
	numpages = {6},
	year = {2019},
	month = {Jan},
	publisher = {American Physical Society},
	doi = {10.1103/PhysRevLett.122.020501},
	url = {https://link.aps.org/doi/10.1103/PhysRevLett.122.020501}
}

@article{Xu2020,
	author  = {Xu, Xiao-Ye and Wang, Qin-Qin and Heyl, Markus and
	Budich, Jan Carl and Pan, Wei-Wei and Chen, Zhe and
	Jan, Munsif and Sun, Kai and Xu, Jin-Shi and
	Han, Yong-Jian and Li, Chuan-Feng and Guo, Guang-Can},
	title   = {Measuring a dynamical topological order parameter in quantum walks},
	journal = {Light: Science \& Applications},
	year    = {2020},
	volume  = {9},
	number  = {1},
	pages   = {7},
	doi     = {10.1038/s41377-019-0237-8},
	url     = {https://doi.org/10.1038/s41377-019-0237-8}
}

@article{PhysRevA.98.022129,
	title = {Dynamical quantum phase transitions in non-Hermitian lattices},
	author = {Zhou, Longwen and Wang, Qing-hai and Wang, Hailong and Gong, Jiangbin},
	journal = {Phys. Rev. A},
	volume = {98},
	issue = {2},
	pages = {022129},
	numpages = {15},
	year = {2018},
	month = {Aug},
	publisher = {American Physical Society},
	doi = {10.1103/PhysRevA.98.022129},
	url = {https://link.aps.org/doi/10.1103/PhysRevA.98.022129}
}

@article{Zhou_2021,
	doi = {10.1088/1367-2630/ac0574},
	url = {https://doi.org/10.1088/1367-2630/ac0574},
	year = {2021},
	month = {jun},
	publisher = {IOP Publishing},
	volume = {23},
	number = {6},
	pages = {063041},
	author = {Zhou, Longwen and Du, Qianqian},
	title = {Non-Hermitian topological phases and dynamical quantum phase transitions: a generic connection},
	journal = {New Journal of Physics},
}

@article{PhysRevB.106.054308,
	title = {Anomaly in the dynamical quantum phase transition in a non-Hermitian system with extended gapless phases},
	author = {Mondal, Debashish and Nag, Tanay},
	journal = {Phys. Rev. B},
	volume = {106},
	issue = {5},
	pages = {054308},
	numpages = {11},
	year = {2022},
	month = {Aug},
	publisher = {American Physical Society},
	doi = {10.1103/PhysRevB.106.054308},
	url ={https://link.aps.org/doi/10.1103/PhysRevB.106.054308}
}

@article{PhysRevB.107.184311,
	title = {Finite-temperature dynamical quantum phase transition in a non-Hermitian system},
	author = {Mondal, Debashish and Nag, Tanay},
	journal = {Phys. Rev. B},
	volume = {107},
	issue = {18},
	pages = {184311},
	numpages = {14},
	year = {2023},
	month = {May},
	publisher = {American Physical Society},
	doi = {10.1103/PhysRevB.107.184311},
	url = {https://link.aps.org/doi/10.1103/PhysRevB.107.184311}
}

@article{Mondal2024,
	author  = {Mondal, Debashish and Nag, Tanay},
	title   = {Persistent anomaly in dynamical quantum phase transition in long-range non-Hermitian {$p$}-wave {Kitaev} chain},
	journal = {The European Physical Journal B},
	year    = {2024},
	volume  = {97},
	number  = {5},
	pages   = {59},
	doi     = {10.1140/epjb/s10051-024-00701-8},
	url     = {https://doi.org/10.1140/epjb/s10051-024-00701-8},
	issn    = {1434-6036}
}

@article{PhysRevLett.132.220402,
	title = {Biorthogonal Dynamical Quantum Phase Transitions in Non-Hermitian Systems},
	author = {Jing, Yecheng and Dong, Jian-Jun and Zhang, Yu-Yu and Hu, Zi-Xiang},
	journal = {Phys. Rev. Lett.},
	volume = {132},
	issue = {22},
	pages = {220402},
	numpages = {7},
	year = {2024},
	month = {May},
	publisher = {American Physical Society},
	doi = {10.1103/PhysRevLett.132.220402},
	url = {https://link.aps.org/doi/10.1103/PhysRevLett.132.220402}
}

@article{92w1-71rt,
	title = {Anatomy of non-Hermitian dynamical quantum phase transitions},
	author = {Fu, Yongxu and Xianlong, Gao},
	journal = {Phys. Rev. B},
	volume = {112},
	issue = {14},
	pages = {L140302},
	numpages = {8},
	year = {2025},
	month = {Oct},
	publisher = {American Physical Society},
	doi = {10.1103/92w1-71rt},
	url = {https://link.aps.org/doi/10.1103/92w1-71rt}
}

@article{Zhang2025SelfNormal,
	author  = {Zhang, Haiting and Wang, Kunkun and Xiao, Lei and Xue, Peng},
	title   = {Self-normal and biorthogonal dynamical quantum phase transitions in non-Hermitian quantum walks},
	journal = {Light: Science \& Applications},
	year    = {2025},
	volume  = {14},
	number  = {1},
	pages   = {253},
	doi     = {10.1038/s41377-025-01919-6},
	url     = {https://doi.org/10.1038/s41377-025-01919-6}
}

@article{f5k4-l93p,
	title = {Nonadiabatic dynamics in non-Hermitian quantum systems from the quantum metric perspective},
	author = {Zhang, Xiang and Hu, Kangyi and Li, Fuxiang},
	journal = {Phys. Rev. B},
	volume = {112},
	issue = {22},
	pages = {224305},
	numpages = {11},
	year = {2025},
	month = {Dec},
	publisher = {American Physical Society},
	doi = {10.1103/f5k4-l93p},
	url = {https://link.aps.org/doi/10.1103/f5k4-l93p}
}

@book{fisher1965nature,
	title={The Nature of Critical Points},
	author={Fisher, M.E.},
	url={https://books.google.com.sg/books?id=cBisnQEACAAJ},
	year={1965},
	publisher={University of Colorado Press}
}

@article{PhysRevB.92.075114,
	title = {Dynamical quantum phase transitions in the Kitaev honeycomb model},
	author = {Schmitt, Markus and Kehrein, Stefan},
	journal = {Phys. Rev. B},
	volume = {92},
	issue = {7},
	pages = {075114},
	numpages = {13},
	year = {2015},
	month = {Aug},
	publisher = {American Physical Society},
	doi = {10.1103/PhysRevB.92.075114},
	url = {https://link.aps.org/doi/10.1103/PhysRevB.92.075114}
}

@article{W1984,
	doi = {10.1088/0305-4470/17/6/026},
	url = {https://doi.org/10.1088/0305-4470/17/6/026},
	year = {1984},
	month = {apr},
	publisher = {},
	volume = {17},
	number = {6},
	pages = {1301},
	author = {W van Saarloos and D A Kurtze},
	title = {Location of zeros in the complex temperature plane: absence of Lee-Yang theorem},
	journal = {Journal of Physics A: Mathematical and General}
}

@article{PhysRevB.110.224302,
	title = {Fisher zeroes and dynamical quantum phase transitions for two- and three-dimensional models},
	author = {Mas{\l{}}owski, Tomasz and Cheraghi, Hadi and Sirker, Jesko and Sedlmayr, Nicholas},
	journal = {Phys. Rev. B},
	volume = {110},
	issue = {22},
	pages = {224302},
	numpages = {14},
	year = {2024},
	month = {Dec},
	publisher = {American Physical Society},
	doi = {10.1103/PhysRevB.110.224302},
	url = {https://link.aps.org/doi/10.1103/PhysRevB.110.224302}
}

@article{PhysRevB.97.060304,
	title = {Dynamical topological invariant after a quantum quench},
	author = {Yang, Chao and Li, Linhu and Chen, Shu},
	journal = {Phys. Rev. B},
	volume = {97},
	issue = {6},
	pages = {060304},
	numpages = {5},
	year = {2018},
	month = {Feb},
	publisher = {American Physical Society},
	doi = {10.1103/PhysRevB.97.060304},
	url = {https://link.aps.org/doi/10.1103/PhysRevB.97.060304}
}

@article{Brody_2014,
	doi = {10.1088/1751-8113/47/3/035305},
	url = {https://doi.org/10.1088/1751-8113/47/3/035305},
	year = {2013},
	month = {dec},
	publisher = {IOP Publishing},
	volume = {47},
	number = {3},
	pages = {035305},
	author = {Brody, Dorje C},
	title = {Biorthogonal quantum mechanics},
	journal = {Journal of Physics A: Mathematical and Theoretical}
}

@article{PhysRevB.109.134301,
	title = {Distribution of Fisher zeros in dynamical quantum phase transitions of two-dimensional topological systems},
	author = {Sacramento, P. D. and Yu, Wing Chi},
	journal = {Phys. Rev. B},
	volume = {109},
	issue = {13},
	pages = {134301},
	numpages = {18},
	year = {2024},
	month = {Apr},
	publisher = {American Physical Society},
	doi = {10.1103/PhysRevB.109.134301},
	url = {https://link.aps.org/doi/10.1103/PhysRevB.109.134301}
}

@article{Fang2012,
	author  = {Fang, Kejie and Yu, Zongfu and Fan, Shanhui},
	title   = {Realizing effective magnetic field for photons by controlling the phase of dynamic modulation},
	journal = {Nature Photonics},
	year    = {2012},
	volume  = {6},
	number  = {11},
	pages   = {782--787},
	doi     = {10.1038/nphoton.2012.236},
	url     = {https://doi.org/10.1038/nphoton.2012.236}
}

@article{Schmidt15,
	author = {M. Schmidt and S. Kessler and V. Peano and O. Painter and F. Marquardt},
	journal = {Optica},
	number = {7},
	pages = {635--641},
	publisher = {Optica Publishing Group},
	title = {Optomechanical creation of magnetic fields for photons on a lattice},
	volume = {2},
	month = {Jul},
	year = {2015},
	url = {https://opg.optica.org/optica/abstract.cfm?URI=optica-2-7-635},
	doi = {10.1364/OPTICA.2.000635}
}

@article{PhysRevA.100.012112,
	title = {Phase-dependent topological interface state and spatial adiabatic passage in a generalized Su-Schrieffer-Heeger model},
	author = {Du, Lei and Wu, Jin-Hui and Artoni, M. and La Rocca, G. C.},
	journal = {Phys. Rev. A},
	volume = {100},
	issue = {1},
	pages = {012112},
	numpages = {9},
	year = {2019},
	month = {Jul},
	publisher = {American Physical Society},
	doi = {10.1103/PhysRevA.100.012112},
	url = {https://link.aps.org/doi/10.1103/PhysRevA.100.012112}
}

@article{Heiss_2012,
	doi = {10.1088/1751-8113/45/44/444016},
	url = {https://doi.org/10.1088/1751-8113/45/44/444016},
	year = {2012},
	month = {oct},
	publisher = {IOP Publishing},
	volume = {45},
	number = {44},
	pages = {444016},
	author = {Heiss, W D},
	title = {The physics of exceptional points},
	journal = {Journal of Physics A: Mathematical and Theoretical}
}

\end{document}